\documentclass[preprint, double-spaced]{revtex4}
\usepackage{wrapfig}
\usepackage{graphicx}
\usepackage{amsmath,bbm}
\usepackage{amssymb,bm}

\begin{document}

\title{Particle production in Ultra-relativistic Heavy-Ion Collisions : A Statistical-Thermal Model Review}
\author{S.~K.~Tiwari\footnote{corresponding author: $sktiwari4bhu@gmail.com$}}
\author{C.~P.~Singh}

\affiliation{Department of Physics, Banaras Hindu University, Varanasi-221005, INDIA}

\begin{abstract}
\noindent
The current status of various thermal and statistical descriptions of particle production in the ultra-relativistic heavy-ion collisions experiments is presented in detail. We discuss the formulation of various types of thermal models of a hot and dense hadron gas (HG) and the methods incorporated in implementing the interactions between hadrons. We first obtain the parameterization of center-of-mass energy ($\sqrt{s_{NN}}$) in terms of temperature ($T$) and baryon chemical potential ($\mu_B$) obtained by analyzing the particle ratios at the freeze-out over a broad energy range from the lowest Alternating Gradient Synchrotron (AGS) energy to the highest Relativistic Heavy-Ion Collider (RHIC) energies. The results of various thermal models together with the experimental results for the various ratios of yields of produced hadrons are then compared. We have derived some new universal conditions emerging at the chemical freeze-out of HG fireball which demonstrate the independence with respect to the energy as well as the structure of the nuclei used in the collision. Further, we perform the calculation of various transport properties of HG such as shear viscosity-to-entropy density ratio ($\eta/s$) etc. using thermal model and compare with the results of other models. We also present the calculation of the rapidity as well as transverse mass spectra of various hadrons in the thermal HG model. The purpose of this review article is to organize and summarize the experimental data obtained in various experiments with heavy-ion collisions and then to examine and analyze them using thermal models so that a firm conclusion regarding the formation of quark-gluon plasma (QGP) can be obtained. 
\\


\end{abstract}
\maketitle 
\section{Introduction}
\noindent
One of the main purpose of various heavy-ion collision programmes running at various places such as Relativistic Heavy-Ion Collider (RHIC) at Brookhaven National Laboratory (BNL) and Large Hadron Collider (LHC) at CERN is to understand the properties of strongly interacting matter and to study the possibility of a phase transition from a confined hot, dense hadron gas (HG) phase to a deconfined and/or chiral symmetric phase of quark matter called as quark-gluon plasma (QGP) \cite{singh:1993,Tannenbaum:2006,Muller:1995,Satz:2000,Satz:2012,Braun:2003,Braun:2009,Braun:2007}. By colliding heavy-ions at ultrarelativistic energies such a phase transition is expected to materialize and QGP can be formed in the laboratory. Unfortunately, the detection of the QGP phase is still regarded as an uphill task. However, the existence of a new form of a matter called as strongly interacting QGP (sQGP) has been demonstrated experimentally \cite{Gyulassy:2005}. There are many compelling evidences $e.\;g.$, elliptic flow, high energy densities and very low viscosity etc. \cite{KAdcox:2005}. However, we do not have supportive evidence that this fluid is associated with the properties quark deconfinement and/or chiral symmetry restoration which are considered as direct indication for QGP formation \cite{KAdcox:2005}. Although various experimental probes have been devised, but a clean unambiguous signal has not yet been outlined in the laboratory. So our prime need at present is to propose some signals to be used for the detection of QGP. However for this purpose, understanding the behaviour and the properties of the background HG is quite essential because if QGP is not formed, matter will continue to exist in the hot and dense HG phase. In the ultra-relativistic nucleus-nucleus collisions, a hot and dense matter is formed over an extended region for a very brief time and it is often called a 'fireball'. The quark matter in the fireball after subsequent expansion and cooling will be finally converted into HG phase. Recently, the studies of the transverse momentum spectra of dileptons \cite{Rapp:2001,Ruuskanen:1987,Rapp:1998,Gallmeister:2000,Karsch:1993,Kajanie:1986,Hatsuda:2002} and hadrons \cite{Heinz:2001,Teaney:2001,Wiedemann:1999,Tomasik:2000} are used to deduce valuable information regarding temperature, energy density of the fireball. The schematic diagram for the conjectured space-time evolution of the fireball formed in the heavy-ion collisions is shown in Fig. 1 \cite{Bjorken:1983}. The space-time evolution consists of four different stages as follows :
($i$) In the initial stage of collisions, labeled as ``Pre-equilibrium'' in Fig. 1, processes of parton-parton hard scatterings may predominantly occur in the overlap region of two colliding nuclei, thus depositing a large amount of energy in the medium. The matter is still not in thermal equilibrium and perturbative QCD models can describe the underlying dynamics as a cascade of freely colliding partons. The time of the pre-equilibrium state is predicted to about $1 \; fm/c$ or less.
($ii$) After the short pre-equilibrium stage, the QGP phase would be formed, in which parton-parton and/or string-string interactions quickly contribute to attain thermal equilibrium in the medium. The energy density of this state is expected to reach above $3-5\; GeV/fm^3$, equivalent to the temperature of $200-300\; MeV$. The volume then rapidly expands and matter cools down.
($iii$) If the first order phase transition is assumed, the ``mixed phase'' is expected to exist between the QGP and hadron phases, in which quarks and gluons are again confined into hadrons at the critical temperature $T_c$. In the mixed phase, the entropy density is transferred into lower degrees of freedom, and therefore, the system is prevented from a fast expansion. This leads to a maximum value in the lifetime of the mixed phase which is expected to last for a relatively long time ($\tau>10\;fm/c$).
($iv$) In the hadronic phase, the system keeps collective expansion via hadron-hadron interactions, decreasing it's temperature. Then, the hadronic interactions freeze after the system reaches a certain size and temperature, and hadrons freely stream out from the medium to be detected. There are two types of freeze-out stages. When inelastic collisions between constituents of the fireball do not occur any longer, we call this as chemical freeze-out stage. Later when the elastic collisions also cease to happen in the fireball, this stage specifies the thermal freeze-out.

\begin{figure}
\includegraphics[height=20em]{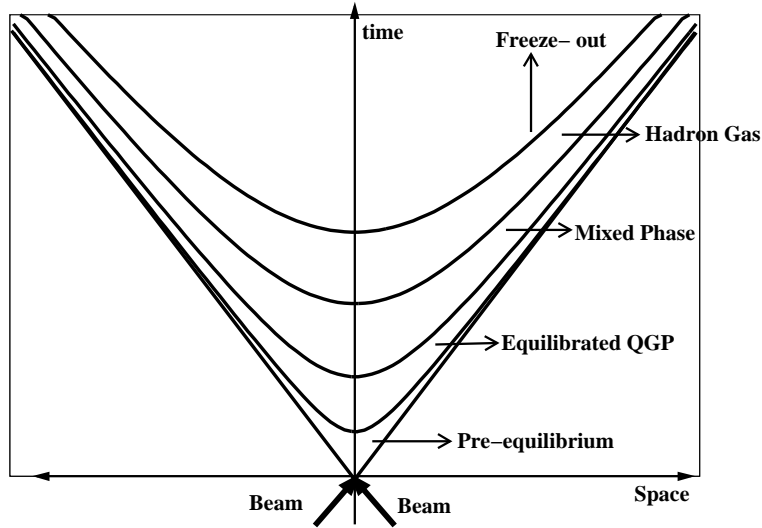}
\caption[]{A schematic diagram of space-time evolution of ultra-relativistic heavy-ion collisions.}
\end{figure}

Since many experiments running at various places measure the multiplicity, ratios etc. of various hadrons, it is necessary to know to which extent the measured hadron yields indicate equilibration. The level of equilibration of particles produced in these experiments is tested by analyzing the particle abundances \cite{Heinz:2001,Braun:1996,Andronic:2006,Tiwari:2012,Mishra:2007,Cleymans:1998,Braun:2001,Magestro:2002,Braun:1995,Braun:1999,Akkelin:2002,Becattini:2001,Keranen:2002,Rafelski:1996,Redlich:2002,Sollfrank:1997,Huovinen:2001,Cleymans:1999,Cleymans1:1999,Cleymans1:1998,Letessier:2000,Braun:2002,Broniowski:2001,Redlich1:2002,Cleymans:1993,Braun:2000,Braun1:2001} or their momentum spectra \cite{Heinz:2001,Teaney:2001,Wiedemann:1999,Tomasik:2000,Braun:1995,Sollfrank:1997,Huovinen:2001,Tiwari:2013} using thermal models. Now, in the first case one establishes the chemical composition of the system, while in the second case, additional information on dynamical evolution and collective flow can be extracted. Furthermore, study of the variations of multiplicity of produced particles with respect to collision energy, the momentum spectra of particles, and ratios of various particles have led to perhaps one of the most remarkable results corresponding to high energy strong interaction physics \cite{Satz:2012}.

Recently various approaches have been proposed for the precise formulation of a proper equation of state (EOS) for hot and dense HG. Lacking lattice QCD results for the EOS at finite baryon density $n_B$, a common approach is to construct a phenomenological EOS for both phases of strongly interacting matter. Among those approaches, thermal models are widely used and indeed are very successful in describing various features of the HG. These models are based on the assumption of thermal equilibrium reached in HG. A simple form of the thermal model of hadrons is the ideal hadron gas (IHG) model in which hadrons and resonances are treated as pointlike and non-interacting particles. The introduction of resonances in the system is expected to account for the existence of attractive interactions among hadrons \cite{Welke:1990}. But in order to account for the realistic behaviour of HG, a short range repulsion must also be introduced. The importance of such correction is more obvious when we calculate the phase transition using IHG picture which shows the reappearance of hadronic phase as a stable configuration in the simple Gibbs construction of the phase equilibrium between the HG and QGP phases at very high baryon densities or chemical potentials. This anomalous situation \cite{Dixit:1983,Cleymans:1986,Cleymans1:1986} cannot be explained because we know that the stable phase at any given $(T,\,\mu_B)$ is the phase which has a larger pressure. Once the system makes a transition to the QGP phase, it is expected to remain in that phase even at extremely large $T$ and $\mu_B$ due to the property of asymptotic freedom of QCD. Moreover, it is expected that the hadronic interactions become significant when hadrons are densely packed in a hot and dense hadron gas. One significant way to handle the repulsive interaction between hadrons is based on a macroscopic description in which the hadrons are given a geometrical size and hence they will experience a hard-core repulsive interaction when they touch each other and consequently a van-der Waals excluded-volume effect is visible. As a result, the particle yields are essentially reduced in comparison to that of IHG model and also the anomalous situation in the phase transition mentioned above disappears. Recently, many phenomenological models incorporating excluded-volume effect have been widely used to account for the properties of hot and dense HG \cite{Cleymans:1987,Cleymans2:1999,Davidson:1991,Kuono:1989,Hagedorn:1980,Rischke:1991,Singh:1996,Panda:2002,Anchishkin:1995,Prasad:2000,Tiwari:1998,Sun:2002}. However, these descriptions usually suffer from some serious shortcomings. First, mostly the descriptions are thermodynamically inconsistent because one does not have a well defined partition function or thermodynamical potential ($\Omega$) from which other thermodynamical quantities can be derived $e.\;g.$ the baryon density ($n_B$)$\neq\partial \Omega/\partial \mu_B$. Secondly, for the dense hadron gas, the excluded-volume model violates causality (i.e., velocity of sound $c_s$ in the medium is greater than the velocity of light). So, although some of the models explain the data very well but such shortcomings make the models mostly unattractive. Sun $et\;al.$ \cite{Sun:2002} have incorporated the effect of relativistic correction in the formulation of an EOS for HG. However, such effect is expected to be very small because the masses of almost all the hadrons present in the hot and dense HG are larger than the temperature of the system, so they are usually treated as non-relativistic particles except pions whose mass is comparable to the temperature but most of the pions come from resonances whose masses are again larger than the temperature of HG \cite{Gorenstein:2012}. In Ref. \cite{Lu:2002}, two-source thermal model of an ideal hadron gas is used to analyze the experimental data on hadron yields and ratios. In this model, the two sources, a central core and a surrounding halo, are in local equilibrium at chemical freeze-out. It has been claimed that the excluded-volume effect becomes less important in the formulation of EOS for hadron gas in the two-source model.

Another important approach used in the formulation of an EOS for the HG phase is mean-field theoretical models \cite{Kapusta:1995,Walecka:1974} and their phenomenological generalizations \cite{Rischke:1988,Zimanyi:1988,Bugaev:1989}. These models use the local renormalizable Lagrangian densities with baryonic and mesonic degrees of freedom for the description of HG phase. These models rigorously respect causality. Most importantly they also reproduce the ground state properties of the nuclear matter in the low-density limit. The short-range repulsive interaction in these models arises mainly due to $\omega$-exchange between a pair of baryons. It leads to the Yukawa potential $V(r)=(G^2/4\,\pi\,r)\,exp(-m_{\omega}\,r)$ , which further gives mean potential energy as $U_B=G^2\,n_B/m_{\omega}$. It means that $U_B$ is proportional to the net baryon density $n_B$. Thus $U_B$ vanishes in the $n_B\rightarrow 0$ limit. In the baryonless limit, hadrons (mesons) can still approach point-like behaviour due to the vanishing of the repulsive interactions between them. It means that in principle one can excite a large number of hadronic resonances at large $T$. This will again make the pressure in the HG phase larger than the pressure in the QGP phase and the hadronic phase would again become stable at sufficiently high temperature and the Gibbs construction can again yield HG phase at large $T$. In some recent approaches this problem has been cured by considering another temperature dependent mean-field $U_{VDW}(n,T)$, where $n$ is the sum of particle and anti-particle number densities. Here $U_{VDW}(n,T)$ represents van-der Waals hard-core repulsive interaction between two particles and depends on the total number density $n$ and is non-zero even when net baryon density $n_B$ is zero in the large temperature limit \cite{Anchishkin:1995,Tiwari:1998}. However, in the high-density limit, the presence of a large number of hyperons and their unknown couplings to the mesonic fields generates a large amount of uncertainty in the EOS of HG in the mean-field approach. Moreover, the assumption of how many particles and their resonances should be incorporated in the system is a crucial one in the formulation of EOS in this approach. The mean-field models can usually handle very few resonances only in the description of HG and hence are not as such reliable \cite{Tiwari:1998}.

In this review, we discuss the formulation of various thermal models existing in literature quite in detail and their applications to the analysis of particle production in ultra-relativistic heavy-ion collisions. We show that it is important to incorporate the interactions between hadrons in a consistent manner while formulating the EOS for hot, dense HG. For repulsive interactions, van-der Waals type of excluded-volume effect is often used in thermal models while resonances are included in the system to account for the attractive interactions. We precisely demand that such interactions must be incorporated in the models in a thermodynamically consistent way. There are still some thermal models in the literature which lack thermodynamical self-consistency. We have proposed a new excluded-volume model where an equal hard-core size is assigned to each type of baryons in the HG while the mesons are treated as pointlike particles. We have successfully used this model in calculating various properties of HG such as number density, energy density etc.. We have compared our results with those of the other models. Further, we have extracted chemical freeze-out parameters in various thermal models by analyzing the particle-ratios over a broad energy range and parameterized them with the center-of-mass energy. We use these parameterizations to calculate the particle-ratios at various center-of-mass energies and compare them with the experimental data. We further provide a proposal in the form of freeze-out conditions for a unified description of chemical freeze-out of hadrons in various thermal models. An extension of the thermal model for the study of the various transport properties of hadrons will also be discussed. We also analyze the rapidity as well as transverse mass spectra of hadrons using our thermal model and examine the role of any flow existing in the medium by matching the theoretical results with the experimental data. Thus the thermal approach indeed provides a very satisfactory description of various features of HG by reproducing a large number of experimental results covering wide energy range from Alternating Gradient Synchrotron (AGS) to Large Hadron Collider (LHC) energy.

\section{Formulation of Thermal Models}
\noindent

Various types of thermal models for HG using excluded-volume correction based on van-der Waals type effect have been proposed. Thermal models have often used the grand canonical ensemble description to write the partition function for the system because it suits well for systems with large number of produced hadrons \cite{Andronic:2006} and/or large volume. However, for nonrelativistic statistical mechanics, the use of a grand canonical ensemble is usually just a matter of convenience \cite{Yen:1997}. Furthermore, the canonical ensemble can be used in case of small systems ($e.\;g.$ peripheral nucleus-nucleus collisions) and for low energies in case of strangeness production \cite{FBecattini:1996} due to canonical suppression of the phase space. Similarly some descriptions also employ isobaric partition function in the derivation of their HG model. We succinctly summarize the features of some models as follows :

\subsection{Hagedorn Model}

In the Hagedorn model \cite{Hagedorn:1980}, it is assumed that the excluded-volume correction is proportional to the pointlike energy density $\epsilon^0$. It is also assumed that the density of states of the finite-size particles in total volume $V$ can be taken as precisely the same as that of pointlike particles in the available volume $\Delta$ where $\Delta=V-\Sigma_i\,V_i^0$ where $V_i^0$ is the eigen volume of the $i^{th}$ particle in the HG. Thus, the grand canonical partition function satisfies the relation :    
\begin{equation}
\begin{array}{lcl}
ln\,Z(T,V,\lambda)=ln\,Z^0(T,\Delta,\lambda).
\end{array}
\end{equation}    
The sum of eigen volumes $\Sigma_i\,V_i^0$ is given by the ratio of the invariant cluster mass to the total energy density and $\lambda$ is the fugacity i.e., $\lambda=exp(\mu/T)$. Hence $\Sigma_i\,V_i^0=E/4\,B=V\,\epsilon/4\,B$ and the energy density $\epsilon=\Delta\,\epsilon^0/V$. $\epsilon^0$ is the energy density when particles are treated as pointlike. Now, using the expression for $\Delta$, one finally gets :
\begin{equation}
\epsilon_i^{ex}=\frac{\epsilon_i^0}{1+\epsilon^0/4\,B}.
\end{equation}                                                                     
When $\epsilon^0/4\,B\gg 1$, $\epsilon=\Sigma_i\,\epsilon_i^{ex}=4\,B$ which is obviously the upper limit for $\epsilon$ since it gives the energy density existing inside a nucleon and usually regarded as the latent heat density required for the phase transition. Here $B$ represents the bag constant. The expressions of number density and pressure can similarly be written :
\begin{equation}
n_i^{ex}=\frac{n_i^0}{1+\epsilon^0/4\,B},
\end{equation}
\begin{equation}
P_i^{ex}=\frac{P_i^0}{1+\epsilon^0/4\,B}.
\end{equation}                      

Here, $n_i^0$ and $P_i^0$ are the number density and pressure of pointlike particles, respectively.

\subsection{Cleymans-Suhonen Model}

In order to include van-der Waals type of repulsion between baryons, Cleymans $et\; al.$ \cite{Cleymans:1987} assigned an equal hard-core radius to each baryon. Consequently, the available volume for baryons is $V-\Sigma_i\,N_iV_i^0$, here $V_i^0$ is the eigen volume of $i^{th}$ baryon and $N_i$ is the total number. As a result, the net excluded number density, pressure and the energy density of a multi-component HG are given as :

\begin{equation}
n^{ex}=\frac{\sum_i\,n_i^0}{1+\sum_i\,n_i^0\,V_i^0},
\end{equation}

\begin{equation}
P^{ex}=\frac{\sum_i\,P_i^0}{1+\sum_i\,n_i^0\,V_i^0},
\end{equation}
                       
\begin{equation}
\epsilon^{ex}=\frac{\sum_i\,\epsilon_i^0}{1+\sum_i\,n_i^0\,V_i^0},
\end{equation}  
where $n_i^0,\; P_i^0$ and $\epsilon_i^0$  are net baryon density, pressure, energy density of pointlike particles, respectively and $\Sigma_i\,n_i^0\,V_i^0$ is the fraction of occupied volume. Kuono and Takagi \cite{Kuono:1989} modified these expressions by considering the existence of a repulsive interaction either between a pair of baryons or between a pair of anti-baryons only. Therefore, the expressions (5), (6) and (7) take the form :
\begin{equation}
n^{ex}=\frac{\sum_i\,n_i^0}{1+\sum_i\,n_i^0\,V_i^0}-\frac{\sum_i\,\bar{n}_{ i}^0}
{1+\sum\,\bar{n}_{ i}^0\,V_i^0}+n_M^0 ,
\end{equation}
\begin{equation}
P^{ex}=\frac{\sum_i\,P_i^0}{1+\sum_i\,n_i^0\,V_i^0}-\frac{\sum_i\,\bar{P}_{i}^0}
{1+\Sigma_i\,\bar{n}_{ i}^0\,V_i^0}+P_M^0,
\end{equation}
\begin{equation}
\epsilon^{ex}=\frac{\sum_i\,\epsilon_i^0}{1+\sum_i\,n_i^0\,V_i^0}+\frac{\sum_i\,\bar{\epsilon}_{i}^0}{1+\sum_i\,\bar{n}_{i}^0\,V_i^0}+\epsilon_M^0,
\end{equation}                                                                             
where $n_i^0$ and $\bar{n}_{i}^0$ are the number density of the pointlike baryons and antibaryons respectively, $\epsilon_i^0(\bar{\epsilon}_{i}^0)$ and $P_i^0(\bar{P}_{i}^0)$ are the corresponding energy density and pressure. Similarly, $n_M^0,\; P_M^0,\;\epsilon_M^0$ are the number density, pressure and energy density of pointlike mesons, respectively.

\subsection{Rischke-Gorenstein-Stocker-Greiner (RGSG) Model}

The above discussed models possess a shortcoming that they are thermodynamically inconsistent because the thermodynamical variables like $n_{B}$ cannot be derived from a partition function or thermodynamical potential ($\Omega$) e.g. $n_B\neq\partial \Omega/\partial \mu_B$. Several proposals have come to correct such type of inconsistencies. Rischke $et\; al.$ \cite{Rischke:1991} have attempted to obtain a thermodynamically consistent formulation. In this model, the grand canonical partition function $Z_G$ for pointlike baryons can be written in terms of canonical partition function $Z_C$ as :  

\begin{equation}
Z_G^0(T,\mu,V)=\sum_{N=0}^{\infty}\,exp(\mu\;N/T)\;Z_C(T,N,V).
\end{equation}
They further modified the canonical partition function $Z_C$ by introducing a step-function in the volume so as to incorporate excluded-volume correction into the formalism. Therefore, the grand canonical partition function (11) finally takes the form :
\begin{equation}
Z_G^{ex}(T,\mu,V-V^0\,N)=\sum_{N=0}^{\infty}\,exp(\mu\,N/T)\,Z_C(T,N,V-V^0\,N)\,\theta(V-V^0\,N).
\end{equation}
The above ansatz is motivated by considering $N$ particles with eigen-volume $V^0$ in a volume $V$ as $N$ pointlike particles in the available volume $V-V^0N$ \cite{Rischke:1991}. But, the problem in the calculation of above Eq. (12) is the dependence of the available volume on the varying number of particles $N$ \cite{Rischke:1991}. To overcome this difficulty one should use Laplace transformation of Eq. (12). Using the Laplace transform, one gets the isobaric partition function as :
\begin{equation}
Z_P=\int_0^{\infty}dV\,exp(-\xi\,V )Z_G^{ex}(T,\mu,V-V^0\,N),
\end{equation}                                                                          
or, :
\begin{equation}
Z_P=\int_0^{\infty}dx\,exp\left\{-x\left[\xi-\frac{ln\,Z_G^0(T,\tilde\mu)}{x}\right]\right\},
\end{equation}                                                                           
where $x=V-V^0\,N$ and $\tilde\mu=\mu-T\,V^0\,\xi$. Finally, we get a transcendental type of equation as follows :
\begin{equation}
P^{ex}(T,\mu)=P^0(T,\tilde\mu),
\end{equation}                                                                     
where,
\begin{equation}
\tilde{\mu}=\mu-V^0\,P^{ex}(T,\mu).
\end{equation}                                                                   
The expressions for the number density, entropy density, and energy density in this model can thus take a familiar form like :
\begin{equation}
n^{ex}(T,\mu)=\frac{\partial P^0(T,\tilde\mu)}{\partial\tilde\mu}\,\frac{\partial\tilde\mu}{\partial\mu}=\frac{n^0(T,\tilde\mu)}{1+V^0\,n^0(T,\tilde\mu)},
\end{equation}
\begin{equation}
s_1^{ex}(T,\mu)=\frac{\partial P^0(T,\tilde\mu)}{\partial T}=\frac{s_1^0(T,\tilde\mu)}{1+V^0\,n^0(T,\tilde\mu)},
\end{equation}
\begin{equation}
\epsilon^{ex}(T,\mu)=\frac{\epsilon^0(T,\tilde\mu)}{1+V^0\,n^0(T,\tilde\mu)}.
\end{equation}                                                                     
These equations resemble with equations (5), and (7) as given in Cleymans-Suhonen model \cite{Cleymans:1987} with $\mu$ replaced by $\tilde\mu$. The above model can be extended for a hadron gas involving several baryonic species as follows :
\begin{equation}
P^{ex}(T,\mu_1,\cdots,\mu_h)=\sum_{i=1}^h\,P_i^0(T,\tilde\mu_i),
\end{equation}                                                                            
where,

\begin{equation}
\tilde{\mu_i}=\mu_i-V_i^0\,P^{ex}(T,\mu_i),
\end{equation}         
with $i=1,\cdots,h$. Particle number density for the $i^{th}$ species can be calculated from following equation :
\begin{equation}
n_i^{ex}(T,\mu_i)=\frac{n_i^0(T,\tilde\mu_i)}{1+\sum_{j=1}^h\,V_j^0\,n_j^0(T,\tilde\mu_j)}.
\end{equation}                                                                            
Unfortunately, the above model involves cumbersome, transcendental expressions which are usually not easy to calculate. Furthermore, this model fails in the limit of $\mu_B=0$ because $\tilde\mu_B$ becomes negative in this limit.

\subsection{New Excluded-Volume Model}

Singh $et\; al.$,~\cite{Singh:1996} have proposed a thermodynamically consistent excluded-volume model in which the grand canonical partition function using Boltzmann approximation can be written as follows :
\begin{equation}
ln\,Z_i^{ex}=\frac{g_i\,\lambda_i}{6\,\pi^2\,T}\,\int_{V_i^0}^{V-\sum_j\,N_j\,V_j^0}
dV\,\int_0^{\infty}\,\frac{k^4\,dk}{\sqrt{k^2+m_i^2}}\,\exp(-\sqrt{k^2+m_i^2}/T),
\end{equation}
where $g_i$ and $\displaystyle\lambda_i = exp(\mu_i/T)$ are the degeneracy factor and the fugacity of $i^{th}$ species of baryons, respectively. $k$ is the magnitude of the momentum of baryons. $V_i^0$ is the eigenvolume assigned to each baryon of $i^{th}$ species and hence $\sum_{j}N_jV_j^0$ becomes the total occupied volume where $N_{j}$ represents the total number of baryons of $j^{th}$ species. We can rewrite Eq. (23) as :
\begin{equation}
ln\,Z_i^{ex}=V(1-\sum_j\,n_j^{ex}\,V_j^0)\,I_i\,\lambda_i,
\end{equation}                                              
where integral $I_i$ is
\begin{equation}
I_i=\frac{g_i}{2\,\pi^2}\left(\frac{m_i}{T}\right)^2\,T^3\,K_2\,(m_i/T).
\end{equation}
Thus we have obtained the grand canonical partition function as given by Eq. (24) by incorporating the excluded volume effect explicitly in the partition function. The number density of baryons after excluded-volume correction ($n_i^{ex}$) can be obtained as :

\begin{equation}
n_i^{ex}=\frac{\lambda_i}{V}\left(\frac{\partial\,\ln\,Z_i^{ex}}{\partial\, \lambda_i}\right)_{T,V}.
\end{equation}

So our prescription is thermodynamically consistent and it leads to a transcendental equation :
\begin{equation}
n_i^{ex}=(1-R)\,\lambda_i\,I_i-I_i\,\lambda_i^2\,\frac{\partial R}{\partial \lambda_i}.
\end{equation}
Here $R=\sum_i\,n_i^{ex}\,V_i^0$ is the fractional occupied volume. It is clear that if we put the factor $\partial R/\partial \lambda_i=0$ and consider only one type of baryons in the system then the above Eq. (27) can be reduced to the thermodynamically inconsistent expression (5). The presence of $\partial R/\partial \lambda_i$ in Eq. (27) thus removes the thermodynamical inconsistency. For single component HG, the solution of Eq. (27) can be taken as :
\begin{equation}
n^{ex}=\frac{1}{V}\,\frac{\int_0^{\lambda}\,d\lambda \exp[-1/I\,V^0\,\lambda]}{\lambda\,exp[-1/I\,V^0\,\lambda]}.
\end{equation}
For a multi-component hadron gas, Eq. (27) takes the following form :
\begin{equation}
R=(1-R)\sum_i I_i\,V_i^0\,\lambda_i-\sum_i I_i\,V_i^0\,\lambda_i^2\,\frac{\partial R}{\partial\lambda_i}.
\end{equation}
Using the method of parametric space~\cite{Uddin}, we write :
\begin{equation}
\lambda_i(t)=\frac{1}{(a_i\,+I_i\,V_i^0\,t)}.
\end{equation}
where $a_i$ is the parameter and $t$ gives the space. We finally get the solution of Eq. (29) as follows :
\begin{equation}
R=1-\frac{\int_t^{\infty}[\exp(-t^{'})/G(t^{'})]dt^{'}}{\exp(-t)/G(t)},
\end{equation}                                                          
where $t$ is a parameter such that :

\begin{equation}
d\lambda_i(t)/dt=-I_i\,\lambda_i^2\,V_i^0,
\end{equation}
and,
\begin{equation}
G(t)=t\,\prod_{i=2}^h(a_i+I_i\,V_i^0\,t).
\end{equation}   
If $\lambda_i$' s and $t$ are known, one can determine $a_i$. The quantity $t$ is fixed by setting $a_1=0$ and one obtains $\displaystyle t=1/I_1\,V_1\,\lambda_1$, here the subscript 1 denotes the nucleon degree of freedom and $h$ is the total number of baryonic species. Hence by using $R$ and $\partial R/\partial\lambda_i$ one can calculate $n_i$. It is obvious that the above solution is not unique. Since it contains some parameters such as $a_i$, one of which has been fixed to zero arbitrarily. Alternatively, one can assume that \cite{Singh:1996} :
\begin{equation}
\frac{\partial R}{\partial\lambda_i}=\frac{\partial \sum_j n_j^{ex}\,V_j^0}{\partial\lambda_i}=\left(\frac{\partial n_i^{ex}}{\partial\lambda_i}\right)\,V_i^0.
\end{equation}               
Here an assumption is made that the number density of $i^{th}$ baryon will only depend on the fugacity of same baryon. Then the Eq. (29) reduces to a simple form as :
\begin{equation}
\frac{\partial n_i^{ex}}{\partial\lambda_i}+n_i^{ex}\left(\frac{1}{I_i\,V_i^0\,\lambda_i^2}+\frac{1}{\lambda_i}\right)=\frac{1}{\lambda_i\,V_i^0}\left(1-\sum_{i\neq j}\,n_j^{ex}\,V_j^0\right).
\end{equation}
The solution of Eq. (35) can be obtained in a straight forward manner as \cite{Singh:1996} :
\begin{equation}
n_i^{ex}=\frac{Q_i(1-\sum_{j\neq i}\,n_j^{ex}V_j^0)}{\lambda_i\,V_i^0}\,
exp\,(1/I_i\,V_i^0\,\lambda_i),
\end{equation}
where
\begin{equation}
Q_i=\int_0^{\lambda_i}\exp(-1/I_i\,V_i^0\,\lambda_i)\,d\lambda_i.
\end{equation}
Now, R can be obtained by using the following relation : 
\begin{equation}
R=\sum_j n_j^{ex}\,V_j^0=\frac{X}{1+X},
\end{equation}
where
\begin{equation}
X=\frac{\sum_i n_i^{ex}\,V_i^0}{1-\sum_i n_i^{ex}\,V_i^0}.
\end{equation}
Here $X$ is the ratio of the occupied to the available volume. Finally, $n_i^{ex}$ can be written as:
\begin{equation}
n_i^{ex}=\frac{(1-R)}{V_i^0}\,\frac{Q_i}{\lambda_i\,exp(-1/I_i\,V_i^0\,\lambda_i)-Q_i}.
\end{equation}
The solution obtained in this model is very simple and easy. There is no arbitrary parameter in this formalism so it can be regarded as a unique solution. However, this theory still depends crucially on the assumption that the number density of $i^{th}$ species is a function of the $\lambda_i$ alone and it is independent of the fugacities of other kinds of baryons. As the interactions between different species become significant in hot and dense HG, this assumption is no longer valid. Moreover, one serious problem crops up, since we cannot do calculation in this model for $T>185$ MeV (and $\mu_B>450$ MeV). This particular limiting value of temperature and baryon chemical potential depends significantly on the masses and the degeneracy factors of the baryonic resonances considered in the calculation.

In order to remove above discrepancies, Mishra $et\;al.$ \cite{Mishra:2007} have proposed a thermodynamically consistent EOS for a hot and dense HG using Boltzmann's statistics. They have proposed an alternative way to solve the transcendental Eq. (27). We have extended this model by using quantum statistics into the grand canonical partition function so that our model works even for the extreme values of temperature and baryon chemical potential. Thus the Eq. (25) can be rewritten as follows \cite{Tiwari:2012}:

\begin{equation}
I_i=\frac{g_i}{6\pi^2 T}\int_0^\infty \frac{k^4 dk}{\sqrt{k^2+m_i^2}} \frac1{\left[exp(\frac{E_i}{T})+\lambda_i\right]},
\end{equation}
and the Eq. (27) takes the following form after using the quantum statistics in the partition function :
\begin{equation}
n_i^{ex} = (1-R)I_i\lambda_i-I_i\lambda_i^2\frac{\partial{R}}{\partial{\lambda_i}}+\lambda_i^2(1-R)I_i^{'}
\end{equation}
where $I_{i}^{'}$ is the partial derivative of $I_{i}$ with respect to $\lambda_{i}$. We can write R in an operator equation form as follows \cite{Mishra:2007,Singh:2009} :
\begin{equation}
R=R_{1}+\hat{\Omega} R
\end{equation}
where $\displaystyle R_{1}=\frac{R^0}{1+R^0}$ with $R^0 = \sum n_i^0V_i^0 + \sum I_i^{'}V_i^0\lambda_i^2$; $n_i^0$ is the density of pointlike baryons of ith species and the operator $\hat{\Omega}$ has the form :
\begin{equation}
\hat{\Omega} = -\frac{1}{1+R^0}\sum_i n_i^0V_i^0\lambda_i\frac{\partial}{\partial{\lambda_i}}.
\end{equation}
Using Neumann iteration method and retaining the series upto $\hat{\Omega}^2$ term, we get
\begin{equation}
R=R_{1}+\hat{\Omega}R_{1} +\hat{\Omega}^{2}R_{1}
\end{equation}
\noindent
After solving Eq. (45), we finally get the expression for total pressure \cite{Singh:1996} of the hadron gas as :
\begin{equation}
P^{ex} = T(1-R)\sum_iI_i\lambda_i + \sum_jP_j^{meson}.
\end{equation}
Here $P_j^{meson}$ is the  pressure due to jth type of meson.

Here we emphasize that we consider the repulsion arising only between a pair of baryons and/or antibaryons because we assign each of them exclusively a hard-core volume. In order to make the calculation simple, we have taken an equal volume $V^{0}=4\pi r^{3}/3$ for each type of baryons with a hard-core radius $r=0.8\;fm$. We have considered in our calculation all baryons and mesons and their resonances having masses upto a cut-off value of $2\;GeV/c^{2}$ and lying in the HG spectrum. Here only those resonances which possess well defined masses and widths have been incorporated in the calculations. Branching ratios for sequential decays have been suitably accounted and in the presence of several decay channels, only dominant mode is included. We have also imposed the condition of strangeness neutrality strictly by putting $\sum_{i}S_{i}(n_{i}^{s}-\bar{n}_{i}^{s})=0$, where $S_{i}$ is the strangeness quantum number of the ith hadron, and $n_{i}^{s}(\bar{n}_{i}^{s})$ is the strange (anti-strange) hadron density, respectively. Using this constraint equation, we get the value of strange chemical potential in terms of $\mu_{B}$. Having done all these things, we proceed to calculate the energy density of each baryon species i by using the following formula :
\begin{equation}
\epsilon_{i}^{ex}=\frac{T^{2}}{V}\frac{\partial lnZ_i^{ex}}{\partial T}+\mu_{i} n_{i}^{ex} 
\end{equation}
Similarly entropy density is :
 \begin{equation}
s=\frac{\epsilon_{i}^{ex}+P^{ex}-\mu_{B}n_{B}-\mu_{S} n_{S}}{T}
\end{equation}

It is evident that this approach is more simple in comparison to other thermodynamically consistent excluded-volume approach which often possesses transcendental final expressions \cite{Rischke:1991,Yen:1997}. Our approach does not involve any arbitrary parameter in the calculation. Moreover, this approach can be used for extremely low as well as extremely large values of $T$ and $\mu_B$  where all other approaches fail to give a satisfying result since we do not use Boltzmann's approximation in our calculation.

\section{Statistical and Thermodynamical Consistency}

Recently, question of statistical and thermodynamical consistency in excluded-volume models for HG has widely been discussed \cite{Gorenstein:2012}. In this section, we re-examine the issue of thermodynamical and statistical consistency of the excluded-volume models. In RGSG model \cite{Rischke:1991}, the single particle grand canonical partition function (12) can be rewritten as follows :

\begin{equation}
Z_G^{ex}(V,T,\mu)= \sum_{N=0}^{\infty}\exp\left(\frac{\mu N}{T}\right)~\frac{(V~-~V^0N)^N}{N!}\,\theta(V-V^0\,N)z^N~
\end{equation}
where,
\begin{equation}
z(T)=\frac{g}{2\pi^2}\int_0^{\infty}k^2dk \exp\left[-~\frac{(k^2+m^2)^{1/2}}{T}\right].
\end{equation}
Here, in this model $N$ in the available volume ($V-V^0N$) is independent of $\mu$. Differentiating Eq. (49) with respect to $\mu$, we get the following Eq. :
\begin{equation}
\frac{\partial Z_G^{ex}}{\partial \mu}=\sum_{N=0}^{\infty} \frac{N}{T} \exp\left(\frac{\mu N}{T}\right)~ \frac{(V~-~V^0N)^N}{N!}\,\theta(V-V^0\,N)z^N~.
\end{equation}
Multiplying both sides of above Eq. (51) by $T/Z_G^{ex}$, we get :
\begin{equation}
\frac{T}{Z_G^{ex}}\frac{\partial Z_G^{ex}}{\partial \mu}=\frac {1}{Z^{ex}}\sum_{N=0}^{\infty} N \exp\left(\frac{\mu N}{T}\right)~ \frac{(V~-~V^0N)^N}{N!}\,\theta(V-V^0\,N)z^N~.
\end{equation}
We know that the expressions for statistical and thermodynamical averages of number of baryons are :
\begin{equation}
\langle N \rangle = \frac {1}{Z_G^{ex}}\sum_{N=0}^{\infty} N \exp\left(\frac{\mu N}{T}\right)~ \frac{(V~-~V^0\overline{N})^N}{N!}~z^N~,
\end{equation}
and
\begin{equation}
\overline{N}~=~T \frac{\partial lnZ_G^{ex}}{\partial \mu}~=~\frac{T}{Z_G^{ex}}\frac{\partial Z_G^{ex}}{\partial \mu},
\end{equation}
respectively. Using Equations (53) and (54) in Eq. (52), we get \cite{Gorenstein:2012} :
\begin{equation}
\overline{N}= \langle N \rangle.
\end{equation}
Thus, we see that in RGSG model, thermodynamical average of number of baryons is exactly equal to the statistical average of number of baryons. Similarly in this model, we can show that :
\begin{equation}
\overline{E}= \langle E \rangle.
\end{equation}

\begin{figure}[t]
\centering
\includegraphics[width=0.5\textwidth]{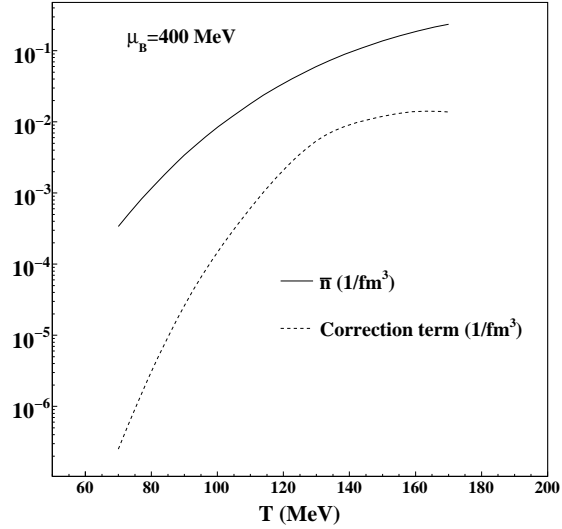}
\caption{Variation of thermodynamical average of the number density of baryons and the ``correction term'' with respect to $T$ at constant $\mu_B$.}
 \label{tdistr}
\end{figure}

Now, we calculate the statistical and thermodynamical averages of number of baryons in our excluded-volume model. The grand canonical partition function in our model ($i.e.$ Eq. (23)) can take the following form :

\begin{equation}
Z^{ex}(V,T,\mu)= \sum_{N=0}^{\infty}\exp\left(\frac{\mu N}{T}\right)~\frac{(V~-~V^0\overline{N})^N}{N!}~z^N~
\end{equation}
where $z$ is given by Eq. (50). We use Boltzmann's statistics for the sake of convenience and consider only one species of baryons. In our model, $\overline{N}$, present in the available volume ($V~-~V^0\overline{N}$), is $\mu$ dependent. However, for multicomponent system one cannot use ``fixed $N$'', because in this case, van-der Waals approximation is not uniquely defined \cite{Bugaev:2008}. So, we use average $N$ in our multicomponent grand partition function. Although, at high temperatures it is not possible to use one component van-der Waals description for a system of various species with different masses \cite{Bugaev:2008}. Now differentiating Eq. (57) with respect to $\mu$, we get :

\begin{equation}
\frac{\partial Z^{ex}}{\partial \mu}=\sum_{N=0}^{\infty} \frac{N}{T} \exp\left(\frac{\mu N}{T}\right)~ \frac{(V~-~V^0\overline{N})^N}{N!}~z^N~\; - \sum_{N=0}^{\infty} \exp\left(\frac{\mu N}{T}\right)~ \frac{(V~-~V^0\overline{N})^{N-1}}{{N-1}!}~z^N V^0 \frac{\partial \overline{N}}{\partial \mu}~.
\end{equation}
Multiplying both sides of above Eq. (58) by $T/Z^{ex}$, we get :

\begin{equation}
\begin{aligned} 
\frac{T}{Z^{ex}}\frac{\partial Z^{ex}}{\partial \mu}=\frac {1}{Z^{ex}}\sum_{N=0}^{\infty} N \exp\left(\frac{\mu N}{T}\right)~ \frac{(V~-~V^0\overline{N})^N}{N!}~z^N~\; - 
\\
\frac {T}{Z^{ex}}\sum_{N=0}^{\infty} \exp\left(\frac{\mu N}{T}\right)~ \frac{(V~-~V^0\overline{N})^{N-1}}{{N-1}!}~z^N V^0 \frac{\partial \overline{N}}{\partial \mu}~.
\end{aligned} 
\end{equation}
Using the definitions (53) and (54), Eq. (59) can take the following form :
\begin{equation}
\overline{N}= \langle N \rangle \; - \frac {T}{Z^{ex}}\sum_{N=0}^{\infty} \exp\left(\frac{\mu N}{T}\right)~ \frac{(V~-~V^0\overline{N})^{N-1}}{{N-1}!}~z^N V^0 \frac{\partial \overline{N}}{\partial \mu}~,
\end{equation}
or,

\begin{equation}
\overline{n}= \langle n \rangle \; - T \langle \frac{n^{0}}{(1-R)} V^0 \frac{\partial \overline{n}}{\partial \mu}~\rangle
\end{equation}
Here $\overline{n}$ is the thermal average of number density of baryons, $n^{0}$ is the number density of pointlike baryons and : 
\begin{equation}
\langle \frac{n^{0}}{(1-R)} V^0 \frac{\partial \overline{n}}{\partial \mu}~\rangle~=~\frac {1}{Z^{ex}}\sum_{N=0}^{\infty} \exp\left(\frac{\mu N}{T}\right)~ \frac{(V~-~V^0\overline{N})^{N}}{{N}!}~z^N \left(\frac {N}{(V~-~V^0\overline{N})} V^0 \frac{\partial \overline{N}}{\partial \mu}\right)~.
\end{equation}
The second term in Eq. (61) is the redundant one and arises because $\overline {N}$, present in the available volume ($V~-~V^0\overline{N}$), is a function of $\mu$. We call this term as 'correction term'. In Fig. 2, we have shown the variation of thermodynamical average of the number density of baryons and the ``correction term'' with respect to $T$ at $\mu_B=400\;MeV$. We see that there is an almost negligible contribution of this ``correction term'' to thermodynamical average of number density of baryons. Although, due to this ``correction term'' the statistical average of the number density of baryons is not exactly equal to it's thermodynamical average but the difference is so small that it can be neglected. Similarly, we can show that such redundant terms appear while calculating statistical average of energy density of baryons and arise due to the temperature dependence of $\overline {N}$. Such terms again give negligible contribution to thermodynamical average of the energy density of baryons. Here, we see that the statistical and thermodynamical averages of physical quantities such as number density, energy density etc. are approximately equal to each other in our model also. Thus, our excluded-volume model is not exactly thermodynamically consistent but it can safely be taken as consistent because the correction term in the averaging procedure appears as negligibly small.

\section{Comparisons Between Model Results and Experimental Data}

In this section, we review various features of hadron gas and present our comparisons between the results of various HG models and the experimental data.

\subsection{Chemical Freeze-out Criteria}

\begin{figure}
\includegraphics[height=25em]{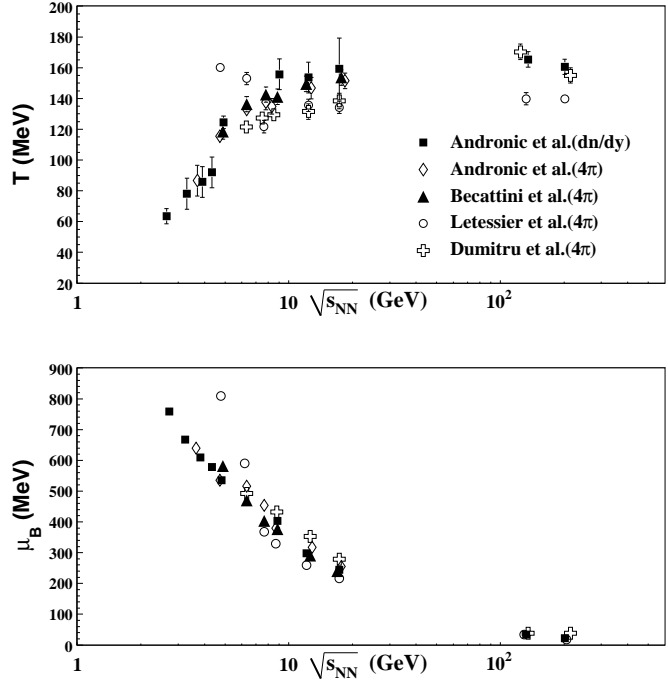}
\caption[]{The energy dependence of chemical freeze-out temperature and baryon chemical potential in various studies. Figure is taken from Ref. \cite{Andronic:2006}.}
\end{figure}

Thermal models provide a systematic study of many important properties of hot and dense HG at chemical freeze-out (where inelastic interactions cease). To establish a relation between chemical freeze-out parameters ($T,\;\mu_B$) and $\sqrt{s_{NN}}$, a common method used is to fit the experimental hadron ratios. Many papers \cite{Andronic:2006,Cleymans1:1999,Cleymans1:1998,Becattini:2004,Letessier:2008,Dumitru:2006} have appeared in which $T$ and $\mu_B$ are extracted in terms of $\sqrt{s_{NN}}$. In Ref. \cite{Andronic:2006}, various hadron ratios are analyzed from $\sqrt{s_{NN}}$=$2.7\; GeV$ to $200\; GeV$ and chemical freeze-out parameters are parameterized in terms of $\sqrt{s_{NN}}$ by using following expressions :
\begin{equation}
T (MeV)=T_{lim}\Big(1-\frac{1}{a+(exp(\sqrt{s_{NN}}(GeV))-b)/c}\Big),
\end{equation}
and,

\begin{equation}
\mu_B=\frac{d}{1+e\sqrt{s_{NN}}}.
\end{equation}  
Here, $a,\;b,\;c,\;d,\;e,$ and $T_{lim}$ (the limiting temperature), are fitting parameters. Various authors \cite{Becattini:2004,Letessier:2008} have included the strangeness suppression factor ($\gamma_s$) in their model while extracting the chemical freeze-out parameters. In the thermal model, $\gamma_s$ is used to account for partial equilibration of the strange particles. Such situation may arise in the elementary $p-p$ collisions and/or peripheral $A-A$ collisions and mostly the use of $\gamma_s$ is warranted in such cases \cite{Becattini:2004,FBecattini:2002}. Moreover, $\gamma_s\approx1$ has been found in the central collisions at RHIC \cite{Kaneta,JAdams:2005}. We do not require $\gamma_s$ in our model as an additional fitting parameter because we assume strangeness is also fully equilibrated in the HG. Also, it has been pointed out that inclusion of $\gamma_s$ in thermal model analysis does not affect the values of fitting parameters $T$ and $\mu_B$ much \cite{Andronic:2006}. Dumitru $et\;al.$ \cite{Dumitru:2006} have used inhomogeneous freeze-out scenario for the extraction of $T$ and $\mu_B$ at various $\sqrt{s_{NN}}$. In a recent paper \cite{Tawfik:2013}, condition of vanishing value of $\kappa\;\sigma^2$ or equivalently $m_4=3 \chi^{2}$ are used to describe the chemical freeze-out line where $\kappa,\;\sigma,\;m_4,$ and $\chi$ are kurtosis, the standard deviation, fourth order moment, and susceptibility, respectively. In Ref. \cite{Gupta:2011}, first time experimental data on $\kappa\;\sigma^2$ and $S\sigma$, here $S$ is skewness, has been compared with the lattice QCD calculations and hadron resonance gas model to determine the critical temperature ($T_c$) for the QCD phase transition. Recently, it is shown that the freeze-out parameters in heavy-ion collisions can be determined by comparing the lattice QCD results for the first three cumulants of net electric charge fluctuations with the experimental data \cite{Bazavov:2012}. In Fig. 3, we have shown the energy dependence of thermal parameters $T$ and $\mu_B$ extracted by various authors. In all the studies, similar behaviour is found except in the Letessier $et\; al.$ \cite{Letessier:2008}, which may be due to usage of many additional free parameters such as light quark occupancy factor ($\gamma_q$), an isospin fugacity etc.. We have also extracted freeze-out parameters by fitting the experimental particle-ratios from the lowest SIS energy to the highest RHIC energy using our model \cite{Tiwari:2012}. For comparison, we have shown the values obtained in other models, $e.\;g.$, IHG model, Cleymans-Suhonen model and RGSG model in Table I. We then parameterize the variables $T$ and $\mu_B$ in terms of $\sqrt{s_{NN}}$ as follows \cite{Cleymans:2006} : 
\begin{equation}
\mu_B=\frac{a}{1+b\sqrt{s_{NN}}}
\end{equation}
\begin{equation}
T=c-d\mu_B^2-e\mu_B^4 
\end{equation}
where the parameters $a$,$b$,$c$,$d$ and $e$ have been determined from the best fit : $a=1.482\pm0.0037$ $GeV$,$ b=0.3517\pm0.009$ ${GeV}^{-1}$, $c=0.163\pm0.0021$ $GeV$,$ d=0.170\pm0.02$ ${GeV}^{-1}$ and $ e=0.015\pm0.01$ ${GeV}^{-3}$. The systematic error of the fits can be estimated via quadratic deviation, $\delta^{2}$ \cite{Andronic:2006} defined as follows :
$$
\delta^{2}=\sum_{i}\frac{(R_i^{exp}-R_i^{therm})^2}{(R_i^{therm})^2},
$$   
where $R_i^{exp}$ and $R_i^{therm}$ are the experimental data and thermal model result of either the hadron yield or the ratio of hadron yields, respectively. 

\begin{figure}
\includegraphics[height=20em]{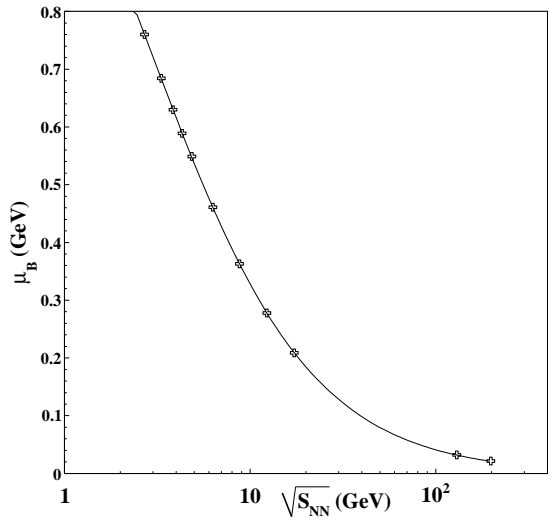}
\caption[]{Variation of baryon chemical potential with respect to centre-of-mass energy in our model.}
\end{figure}

\begin{figure}
\includegraphics[height=20em]{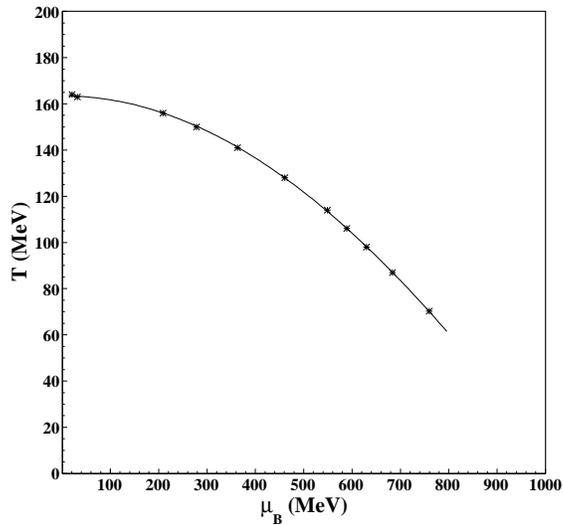}
\caption[]{Variation of chemical freeze-out temperature with respect to baryon chemical potential in our model.}
\end{figure}

\begin{wraptable}{c}{15 cm} 
\caption{Thermal parameters $(T,\;\mu_B)$ values obtained by fitting the experimental particle-ratios in different model calculations.}
\begin{tabular}{l|l|l|l|l|l|l|l|l|l|l|l|l}
\hline
 \boldmath{$\sqrt{s_{NN}}$}\textbf{(GeV)} & \multicolumn{3}{l|}{\textbf{IHG Model}}&\multicolumn{3}{l|}{\textbf{RGSG Model}} &\multicolumn{3}{l|}{\textbf{Cleymans-Suhonen Model}}  &\multicolumn{3}{l}{\textbf{Our Model}} \\
\cline{2-13}
 &\textbf{T}&\boldsymbol{$\mu_{B}$}&\boldsymbol{$\delta^{2}$}&\textbf{T}&\boldsymbol{$\mu_{B}$}&\boldsymbol{$\delta^{2}$}&\textbf{T}&\boldsymbol{$\mu_{B}$}&\boldsymbol{$\delta^{2}$}&\textbf{T}&\boldsymbol{$\mu_{B}$}&\boldsymbol{$\delta^{2}$}\\
\hline\hline

2.70  & 60     &740      & 0.85       & 60      & 740    & 0.75           & 70     & 753        & 1.19       & 70     & 760        & 1.15\\
3.32  & 80     &670      & 0.89       & 78      & 680    & 0.34           & 89     & 686        & 0.75       & 90     & 670        & 0.45\\
3.84  & 100    &645      & 0.50       & 86      & 640    & 0.90           & 101    & 639        & 0.37       & 100    & 640        & 0.34\\
4.32  & 101    &590      & 0.70       & 100     & 590    & 0.98           & 109    & 600        & 0.17       & 105    & 600        & 0.23\\
8.76  & 140    &380      & 0.45       & 145     & 406    & 0.62           & 144    & 386        & 0.05       & 140    & 360        & 0.25\\
12.3  & 148    &300      & 0.31       & 150     & 298    & 0.71           & 153    & 300        & 0.03       & 150    & 276        & 0.20\\
17.3  & 160    &255      & 0.25       & 160     & 240    & 0.62           & 158.6  & 228        & 0.63       & 155    & 206        & 0.27\\
130   & 172.3  &35.53    & 0.10       & 165.5   & 38     & 0.54           & 165.8  & 35.84      & 0.15       & 163.5  & 32         & 0.05\\
200   & 172.3  & 23.53   & 0.065      & 165.5   & 25     & 0.60           & 165.9  & 23.5       & 0.10       & 164    & 20         & 0.05\\ \hline

\hline

\end{tabular}

\end{wraptable}

In this analysis, we have used full phase space (4$\pi$) data at all center-of-mass energies except at RHIC energies where only mid-rapidity data are available for all the ratios. Moreover, the midrapidity and full phase space data at these energies differ only slightly as pointed out by C. Alt $et\; al.$ for $K^+/\pi^+$ and $K^-/\pi^-$ ratios \cite{Alt:2008}. In Fig. 4, we have shown the parametrization of the freeze-out values of baryon chemical potential with respect to $\sqrt{s_{NN}}$ and similarly in Fig. 5, we have shown the chemical freeze-out curve between temperature and baryon chemical potential \cite{Tiwari:2012}.

\subsection{Hadron ratios}

\begin{figure}
\includegraphics[height=20em]{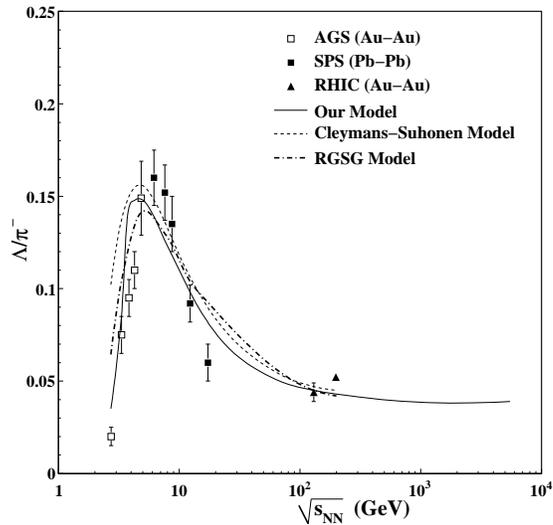}
\caption[]{The energy dependence of the $\Lambda/\pi^-$ ratio. Lines are the results of various thermal models \cite{Tiwari:2012,Cleymans:1987,Rischke:1991}. Points are the experimental data \cite{Alt:2008,Alt1:2008,Aggarwal:2011,Andronic:2010,Abelev:2012}. RHIC data are at midrapidity.}
\end{figure}

In an experimental measurement of various particle ratios at various centre-of-mass energies \cite{Gazdzicki:2004,Afanasiev:2002,Anticic:2004,Ahle:1998,Albergo:2002}, it is found that there is an unusual sharp variation in the $\Lambda/\pi^{-}$ ratio increasing upto the peak value. This strong variation of the ratio with energy indicates the critical temperature of QCD phase transition \cite{Chatterjee:2010} between HG and QGP \cite{Stock:2004,Noronha:2010} and a nontrivial information about the critical temperature $T_C\approx176\;MeV$ has been extracted \cite{Noronha:2010}. Fig. 6 shows the variation of $\Lambda/\pi^{-}$ with $\sqrt{s_{NN}}$. We compare the experimental data with various thermal models \cite{Tiwari:2012,Cleymans:1987,Rischke:1991} and find that our model calculation gives much better fit to the experimental data in comparison to other models. We get a sharp peak around centre-of-mass energy of 5 GeV and our results thus almost reproduce all the features of the experimental data.

\begin{figure}
\includegraphics[height=20em]{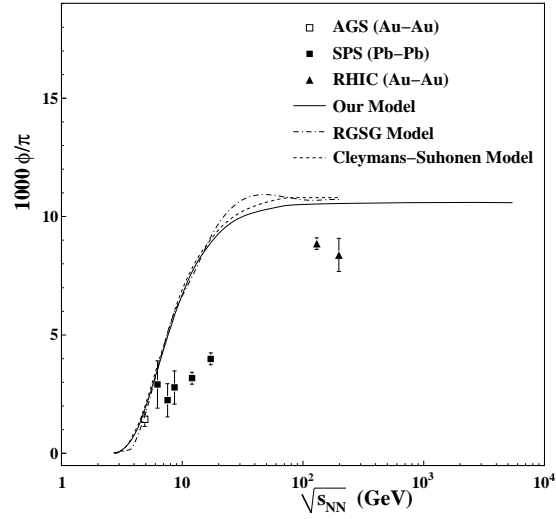}
\caption[]{The energy dependence of $\phi/\pi$ ratio. Lines are the results of various thermal models \cite{Tiwari:2012,Cleymans:1987,Rischke:1991}. Points are the experimental data \cite{Alt:2008,Alt1:2008,Aggarwal:2011,Andronic:2010,Abelev:2012}. RHIC data are at midrapidity. }
\end{figure}

\begin{figure}
\includegraphics[height=25em]{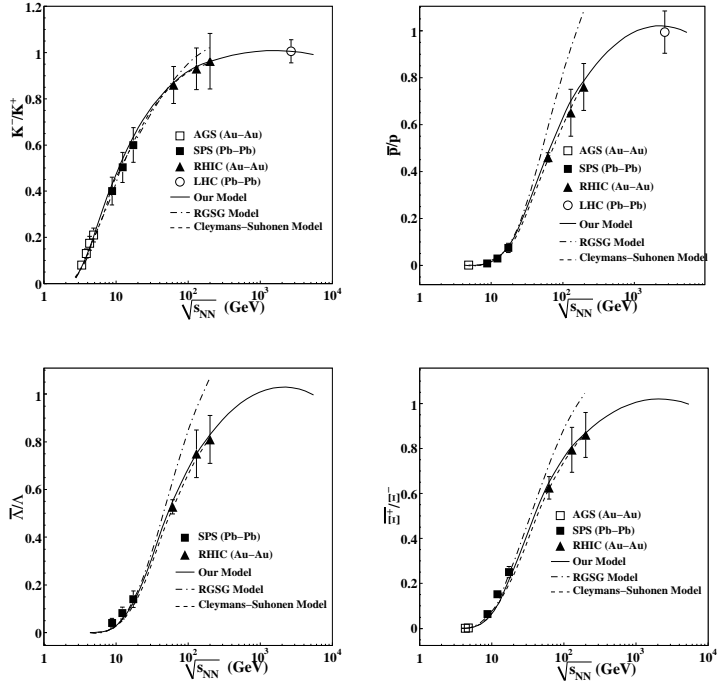}
\caption[]{The energy dependence of anti hadron to hadron ratios. Points are the experimental data \cite{Alt:2008,Alt1:2008,Aggarwal:2011,Andronic:2010,Abelev:2012} and lines are results of various models. RHIC and LHC data are at midrapidity.}
\end{figure}

In Fig. 7, we have shown the variations of $\phi/\pi$ ratio with $\sqrt{s_{NN}}$. The $\phi$ yields in the thermal models are often much higher in comparison to data. We notice that no thermal model can suitably account for the multiplicity-ratio of multi strange particle since $\phi$ is $s\bar{s}$ hidden-strange quark combination. However, quark coalescence model assuming a QGP formation has claimed to explain the results \cite{Molnar:2003} successfully. In the thermal models, the results for the multistrange particles raises doubt over the degree of chemical equilibration for strangeness reached in the HG fireball. We can use an arbitrary parameter $\gamma_s$ as used in several models. The failures of thermal models in these cases may indicate the presence of QGP formation but it is still not clear. In Fig. 8, we have shown the energy dependence of antiparticle to particle ratios e. g. $K^-/K^+$, $\bar{p}/p$, $\bar{\Lambda}/\Lambda$, and $\bar{\Xi^{+}}/\Xi^{-}$. These ratios increase sharply with respect to $\sqrt{s_{NN}}$ and then almost saturate at higher energies reaching the value equal to $1.0$ at LHC energy. On comparison with the experimental data we find that almost all the thermal models describe these data successfully over all center-of-mass energies. However, RGSG model \cite{Rischke:1991} fails to describe the data at SPS and RHIC energies in comparison to other models \cite{Tiwari:2012}.

\subsection{Thermodynamical Properties}

\begin{figure}
\includegraphics[height=20em]{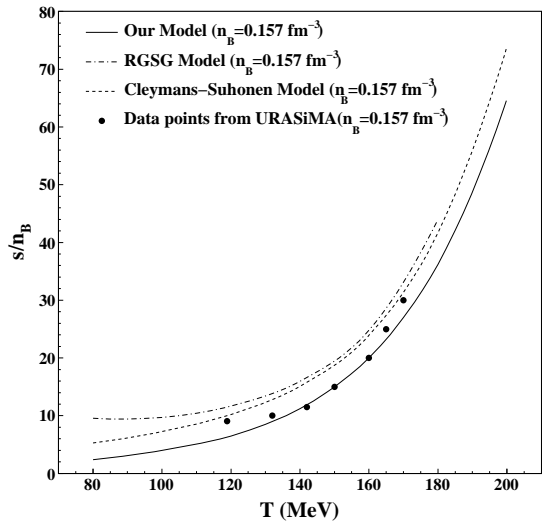}
\caption[]{Variation of $s/n_B$ with respect to temperature at constant net baryon density. Lines show the results of various thermal models and points are the data calculated by Sasaki using URASiMA event generator.}
\end{figure}

We present the thermal model calculations of various thermodynamical properties of HG such as entropy per baryon ($s/n_B$) and energy density etc. and compare the results with the predictions of a microscopic model URASiMA event generator developed by Sasaki \cite{Sasaki:2001}. URASiMA (ultra-relativistic AA collision simulator based on multiple scattering algorithm) is a microscopic model which includes the realistic interactions between hadrons. In URASiMA event generator, molecular-dynamical simulations for a system of a HG are performed. URASiMA includes the multibody absorptions, which are the reverse processes of multiparticle production and are not included in any other model. Although, URASiMA gives a realistic EOS for hot and dense HG, it does not include antibaryons and strange particles in their simulation, which is very crucial. In Fig. 9, we have plotted the variation of $s/n_B$ with respect to temperature ($T$) at fixed net baryon density ($n_B$). $s/n_B$ calculated in our model shows a good agreement with the results of Sasaki \cite{Sasaki:2001} in comparison to other excluded-volume models. It is found that thermal model approach, which incorporates macroscopic geometrical features gives a close results with the simulation involving microscopic interactions between hadrons. There are various parameters such as coupling constants of hadrons etc. appear in URASiMA model due to interactions between hadrons. It is certainly encouraging to find an excellent agreement between the results obtained with two widely different approaches. 

\begin{figure}
\includegraphics[height=20em]{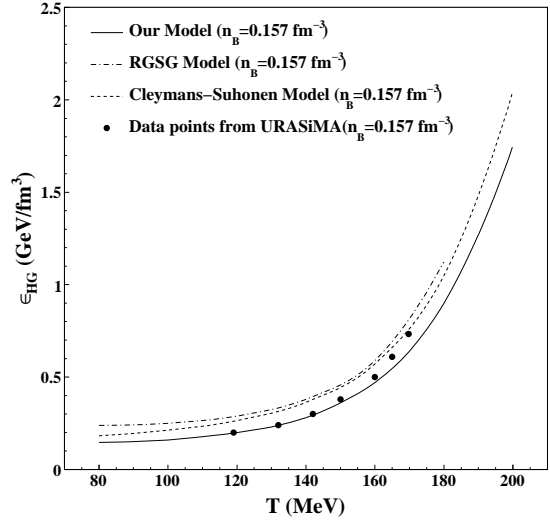}
\caption[]{Variation of energy density with respect to temperature at constant net baryon density. Lines are the results of various thermal models and points are the data calculated by Sasaki using URASiMA event generator.}
\end{figure}

Fig. 10 represents the variation of the energy density of HG with respect to $T$ at constant $n_B$. Again our model calculation is more closer to the result of URASiMA in comparison to other excluded-volume models. Energy density increases very slowly with the temperature initially and then rapidly increases at higher temperatures.

\subsection{Causality}

\begin{figure}
\includegraphics[height=20em]{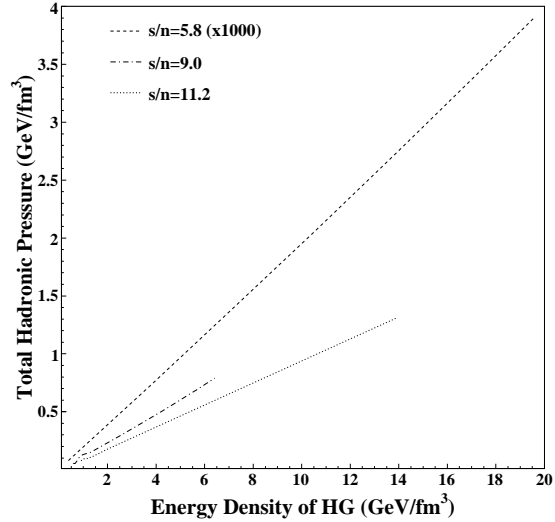}
\caption[]{Variations of total hadronic pressure with respect to energy density of HG at fixed entropy per particle $s/n$. Our calculations show linear relationship and slope of the lines give square of the velocity of sound ${c_{s}}^{2}$ \cite{Tiwari:2012}.}
\end{figure}

\begin{figure}
\includegraphics[height=20em]{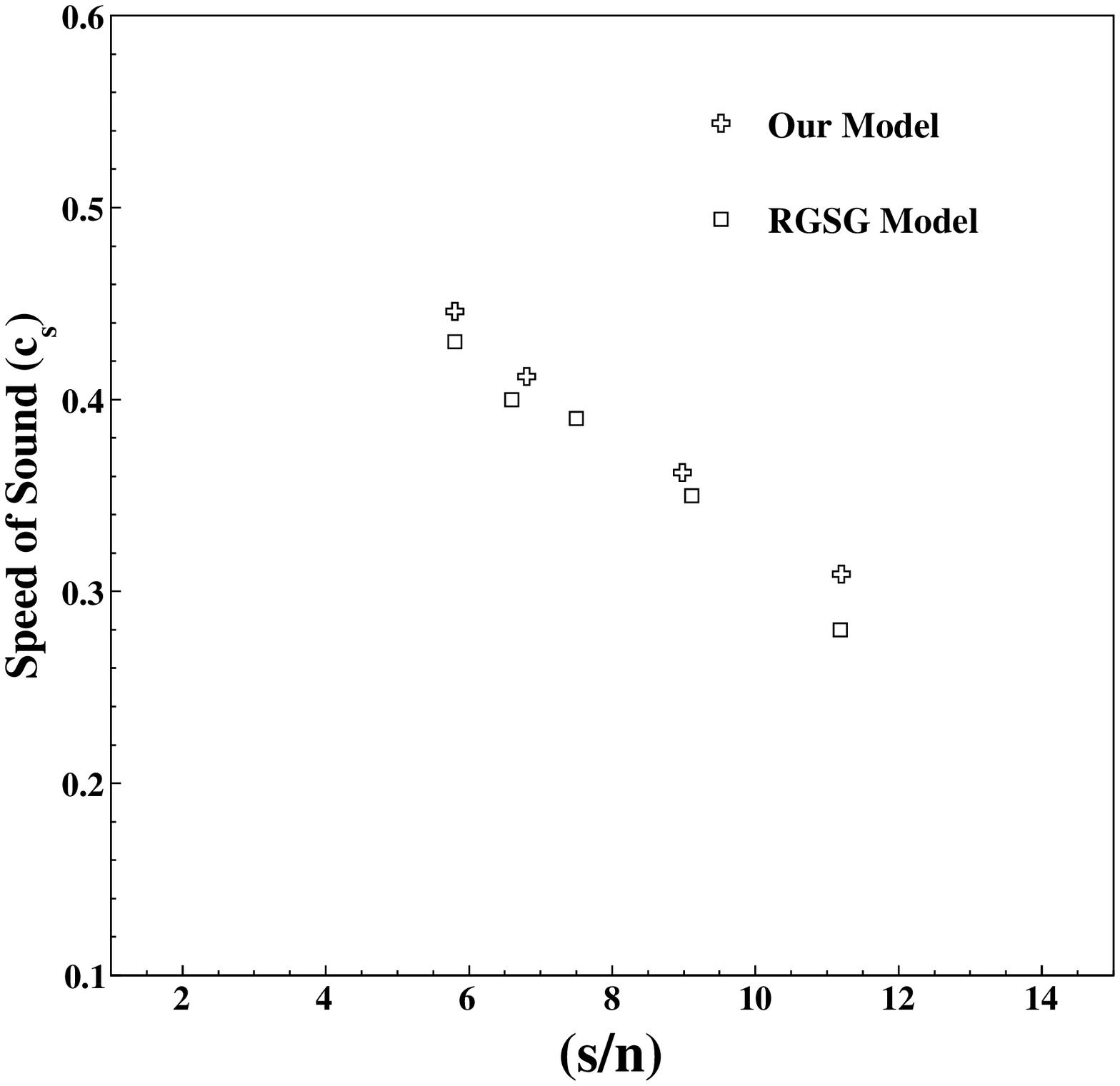}
\caption[]{Variation of velocity of sound in the hot, dense HG medium with respect to entropy per particle ($s/n$).}
\end{figure}

One of the deficiencies of excluded-volume models is the violation of causality in the hot and dense hadron gas i.e., the sound velocity $c_s$ is larger than the velocity of light $c$ in the medium. In other words, $c_s>1$ in the unit of $c=1$, means that the medium transmits information at a speed faster than $c$ \cite{Prasad:2000}. Since, in this article we are discussing the results of various excluded-volume models, it would be interesting to see whether these models respect causality or not. In Fig. 11, we have plotted the variations of the total hadronic pressure $P$ as a function of the energy density $\epsilon$ of the HG at a fixed entropy per particle using our model calculation \cite{Tiwari:2012}. We find for a fixed $s/n$, the pressure varies linearly with respect to energy density. In Fig. 12, we have shown the variation of $c_s$ (${c_s}^2=\partial P/\partial\epsilon$ at fixed $s/n$) with respect to $s/n$. We find that $c_s\leq0.58$ in our model with interacting particles. We get $c_s=0.58$ (i.e. $1/\sqrt{3}$) for an ideal gas consisting of ultra-relativistic particles. This feature endorses our viewpoint that our model is not only thermodynamically consistent but it does not involve any violation of causality even at large density. Similarly in RGSG model \cite{Rischke:1991}, we do not notice that the value of $c_s$ exceeds $1$ as shown in Fig. 12. It should be mentioned that we are using full quantum statistics in all the models taken for comparisons here. However, we find that the values in the RGSG model cannot be extracted when temperature of the HG exceeds $250\;MeV$. No such restriction applies for our model.

\subsection{Universal freeze-out criteria}

\begin{figure}
\includegraphics[height=20em]{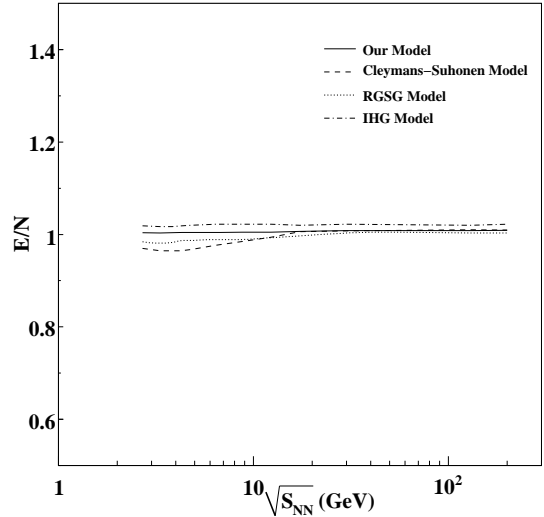}
\caption[]{Variation of $E/N$ with $\sqrt{s_{NN}}$. IHG model calculation is shown by dash dotted line,Cleymans-Suhonen and RGSG model calculations are shown by dashed and dotted line, respectively. Solid line shows the result of our model.}
\end{figure}

One of the most remarkable success of thermal models is in explaining the multiplicities and the particle ratios of various particles produced in heavy-ion experiments from the lowest SIS energy to maximum LHC energy. Some properties of thermal fireball are found to be common to all collision energies which give a universal freeze-out conditions in heavy-ion collisions. Now, we review the applicability of thermal models in deriving some useful chemical freeze-out criteria for the fireball. Recent studies \cite{Braun:1999,Cleymans:1999,Cleymans:2006,Braun1:2002,Magas:2003,Tawfik:2005} predict that following empirical conditions to be valid on the entire freeze-out hypersurface of the fireball : $(i)$ energy per hadron always has a fixed value at $1.08\;GeV$, $(ii)$ sum of baryon and anti-baryon densities $n_{B}+n_{\bar{B}}=0.12/{fm}^3$, $(iii)$ normalized entropy density $s/T^{3}\approx7$. Further, Cleymans $et\; al.$ \cite{Cleymans:2006} have found that all the above conditions separately give a satisfactory description of the chemical freeze-out parameters $T$ and $\mu_B$ in an IHG picture only. Moreover, it was also found that these conditions are independent of collision energy and the geometry of colliding nuclei. Furthermore, Cleymans $et\; al.$ \cite{Cleymans:2006} have hinted that incorporation of excluded-volume correction leads to wild as well as disastrous effects on these conditions. The purpose in this section is to reinvestigate the validity of these freeze-out criteria in excluded-volume models. Along with these conditions, a condition, formulated by using percolation theory is also proposed as a chemical freeze-out condition \cite{Magas:2003}. An assumption is made that in the baryonless region the hadronic matter freezes-out due to hadron resonances and vacuum percolation, while in the baryon rich region the freeze-out takes place due to baryon percolation. Thus the condition which describes the chemical freeze-out line is formulated by following equation \cite{Magas:2003} :  
\begin{equation}
n(T,\mu)=\frac{1.24}{V_h}\Big[1-\frac{n_B(T,\mu)}{n(T,\mu)}\Big]+\frac{0.34}{V_h}\frac{n_B(T,\mu)}{n(T,\mu)},
\end{equation}
where $V_h$ is the volume of a hadron. The numbers 1.24 and 0.34 are obtained within percolation theory \cite{Isichenko:1992}.

\begin{figure}
\includegraphics[height=20em]{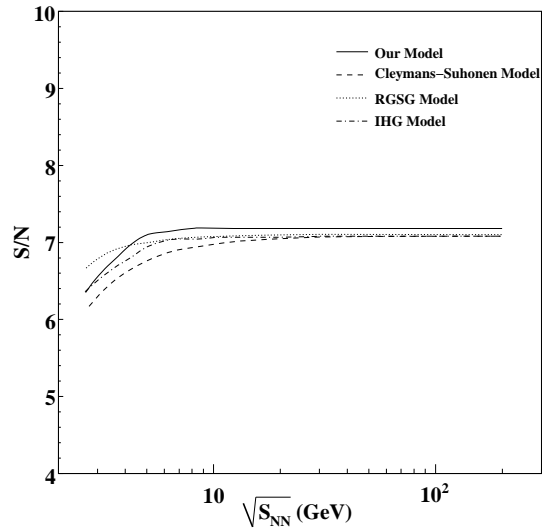}
\caption[]{Variation of $S/N$ with $\sqrt{s_{NN}}$. IHG model calculation is shown by dash dotted line,Cleymans-Suhonen and RGSG model calculation shown by dashed and dotted line respectively. Solid line shows the calculation by our model}
\end{figure}

In Fig. 13, we have shown the variation of $E/N$ with respect to $\sqrt{s_{NN}}$ at the chemical freeze-out point of the fireball. The ratio $E/N$ shows a constant value of 1.0 in our model and it shows also a remarkable energy independence. Similarly the curve in IHG model shows that the value for $E/N$ is slightly larger than one as reported in \cite{Cleymans:2006}. However, results support the finding that $E/N$ is almost independent of energy and also of the geometry of the nuclei. Most importantly we notice that the inclusion of the excluded-volume correction does not change the result much which is contrary to the claim of Cleymans $et\; al.$ \cite{Cleymans:2006}. The condition $E/N\approx1.0 GeV$ was successfully used in the literature to make predictions \cite{Redlich:2002} of freeze-out parameters at SPS energies of 40 and 80 A GeV for Pb-Pb collisions long before the data were taken \cite{Tiwari:2012}. Moreover, we have also shown in Fig. 13, the curves in the Cleymans-Suhonen model \cite{Cleymans:1987} and the RGSG model \cite{Rischke:1991} and we notice a small variation with $\sqrt{s_{NN}}$ particularly at lower energies. In Fig. 14, we study a possible new freeze-out criterion which was not proposed earlier. We show that the quantity entropy per particle i.e. $S/N$ yields a remarkable energy independence in our model calculation. The quantity $S/N\approx7.0$ describes the chemical freeze-out criteria and is almost independent of the centre-of-mass energy in our model calculation. However, the results below $\sqrt{s_{NN}}=6\;GeV$ do not give promising support to our criterion and reveal some energy-dependence also. This criterion thus indicates that the possible use of excluded-volume models and the thermal descriptions at very low energies is not valid for the HG. Similar results were obtained in the RGSG, Cleymans-Suhonen and IHG model also \cite{Tiwari:2012}. The conditions $i. e.$ $E/N\approx1.0 GeV$ and $S/N\approx7.0$ at the chemical freeze-out form a constant hypersurface from where all the particles freeze-out and all kinds of inelastic collisions cease simultaneously and fly towards the detectors. Thus all particles attain thermal equilibrium at the line of chemical freeze-out and when they come out from the fireball they have an almost constant energy per particle ($\approx1.0$) and entropy per particle ($\approx7.0$). Moreover, these values are independent of the initial collision energy as well as the geometry of the colliding nuclei.

Our finding lends support to the crucial assumption of HG fireball achieving chemical equilibrium in the heavy-ion collisions from the lowest SIS to RHIC energy and the EOS of the HG developed by us indeed gives a proper description of the hot and dense fireball and its subsequent expansion. However, we still do not get any information regarding QGP formation from these studies. The chemical equilibrium once attained by the hot and dense HG removes any memory regarding QGP existing in the fireball before HG phase. Furthermore, in a heavy-ion collision, a large amount of kinetic energy becomes available and part of it is always lost during the collision due to dissipative processes. In thermal description of the fireball, we ignore the effect of such processes and we assume that all available kinetic energy (or momentum) is globally thermalized at the freeze-out density. Experimental configuration of the collective flow developed in the hot, dense matter reveals the unsatisfactory nature of the above assumption.

\subsection{Transport Properties of HG}

Transport coefficients are very important tools in quantifying the properties of strongly interacting relativistic fluid and its critical phenomena i.e., phase transition, critical point etc. \cite{Kapusta:1981,Polyakov:1969,Karsch:2008}. The fluctuations cause the system to depart from equilibrium and a non-equilibrated system is created for a brief time. The response of the system to such fluctuations is essentially described by the transport coefficients e.g., shear viscosity, bulk viscosity, speed of sound etc. Recently the data for the collective flow obtained from RHIC and LHC experiments indicate that the system created in these experiments behaves as strongly interacting perfect fluid \cite{Heinz:2002}, whereas we expected that QGP created in these experiments should behave like a perfect gas. The perfect fluid created after the phase transition indicate a very low value of shear viscosity to entropy ratio so that the dissipative effects are negligible and the collective flow is large as obtained by heavy ion collision experiments \cite{Gyulassy:2005,Shuryak:2005,Romatschke:2007}. There were several analytic calculations for $\eta$ and $\eta/s$ of simple hadronic systems \cite{Gavin:1985,Dobado:2002,Prakash:1993,Chen:2007,Chen1:2007,Itakura:2008,Khvorostukhin:2010} along with some sophisticated microscopic transport model calculations \cite{Muronga:2004,Muroya:2005,Demir:2009} in the literature. Furthermore, some calculations predict that the minimum of shear viscosity to entropy density is related with the QCD phase transition \cite{Csernai:2006,Noronha:2009,Lacey:2007,Dobado:2009,Pal:2010}. Similarly sound velocity is an important property of the matter created in heavy ion collision experiments because the hydrodynamic evolution of this matter strongly depends on it. A minimum in the sound-velocity has also been interpreted in terms of a phase transition \cite{Braun:1996,Noronha:2009,Castorina:2010,Cleymans:2011,Gavai:1986,Redlich:1986,Karsch:2007,Prorok} and further, the presence of a shallow minimum corresponds to a cross-over transition \cite{Chojnacki:2007}. In view of the above, it is worthwhile to study in detail the transport properties of the HG in order to fully comprehend the nature of the matter created in the heavy-ion collisions as well as the involved phase transition phenomenon. In this section, we have used thermal models to calculate the transport properties of HG such as shear viscosity to entropy ratio etc. \cite{Tiwari:2012}.  

We calculate the shear viscosity in our thermal model as was done previously by Gorenstein $et\; al.$ \cite{Gorenstein:2008} using RGSG model. According to molecular kinetic theory, we can write the dependence of the shear viscosity as follows \cite{Lifschitz} :
\begin{equation}
\eta \propto n\;l\;\langle|{\bf p}|\rangle , 
\end{equation}
where $n$ is the particle density, $l$ is the mean free path, and $\bf {p}$ is the average thermal momentum of the baryons or antibaryons. For the mixture of particle species with different masses and with the same hard-core radius $r$, the shear viscosity can be calculated by the following equation \cite{Gorenstein:2008} :
\begin{equation}
\eta=\frac{5}{64 \sqrt{8} \;r^2}\sum_{i}\langle|{\bf{p}_{i}}|\rangle\times \frac{n_{i}}{n},
\end{equation}
where $n_{i}$ is the number density of the ith species of baryons (anti-baryons) and $n$ is the total baryon density. In Fig. 15, we have shown the variation of $\eta/s$ with respect to temperature as obtained in our model for HG having a baryonic hard-core size $r=0.5$ fm, and compared our results with those of Gorenstein $et\; al.$ \cite{Gorenstein:2008}. We find that near the expected QCD phase transition temperature ($T_{c}=170-180$ MeV), $\eta/s$ shows a lower value in our HG model than the value in other model. In fact, $\eta/s$  in our model looks close to the lower bound ($1/4 \pi)$ suggested by AdS/QCD theories \cite{Policastro:2001}. Recently, measurements in Pb-Pb collisions at the Large Hadron Collider (LHC) support the value $\displaystyle\eta/s\approx 1/4 \pi$ when compared with the viscous fluid hydrodynamic flow \cite{Qiu:2012}.

\begin{figure}
\begin{center}
\includegraphics[height=15em]{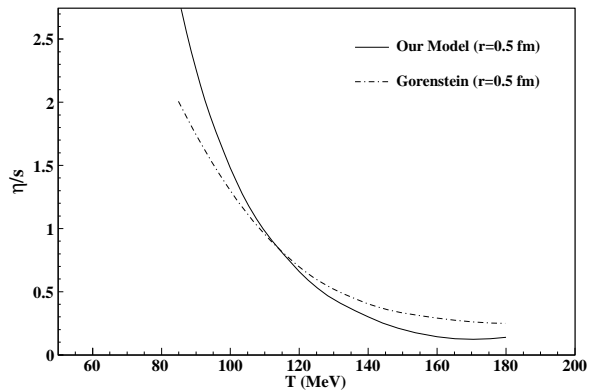}
\caption[]{Variation of $\eta/s$ with temperature for $\mu_{B}=0$ in our model and a comparison with the results obtained by Gorenstein $et\; al.$ \cite{Gorenstein:2008}.}
\end{center}
\end{figure}
\begin{figure}
\begin{center}
\includegraphics[height=15em]{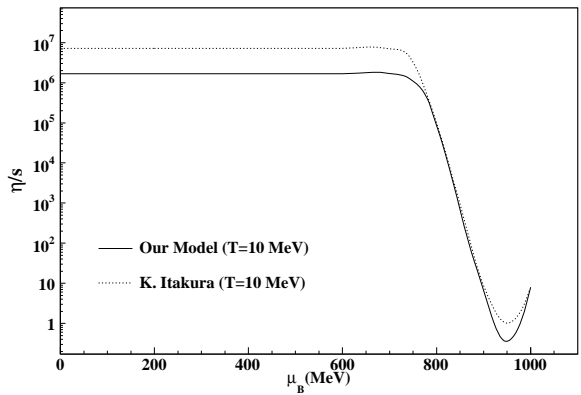}
\caption[]{Variation of $\eta/s$ with  respect to baryon chemical potential ($\mu_{B}$) at very low temperature 10 MeV. Solid line represents our calculation \cite{Tiwari:2012} and dotted curve is by K. Itakura $et\; al.$ \cite{Itakura:2008}.}
\end{center}
\end{figure}
\begin{figure}
\begin{center}
\includegraphics[height=15em]{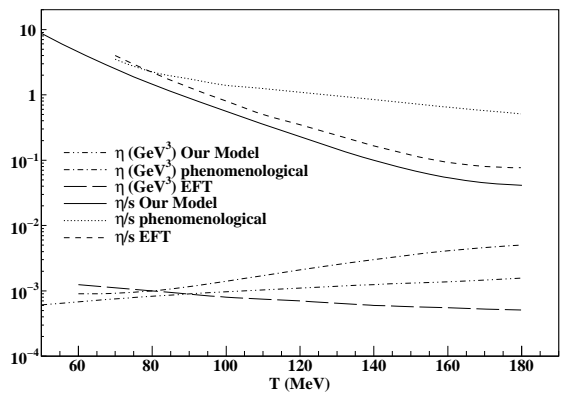}
\caption[]{Variation of $\eta$ in unit of $(GeV)^3$ and $\eta/s$ with respect to temperature at $\mu_{B}=300 MeV$ in our model \cite{Tiwari:2012} and a comparison with the results obtained by K. Itakura $et\; al.$ \cite{Itakura:2008}.}
\end{center}
\end{figure}

\begin{figure}
\begin{center}
\includegraphics[height=15em]{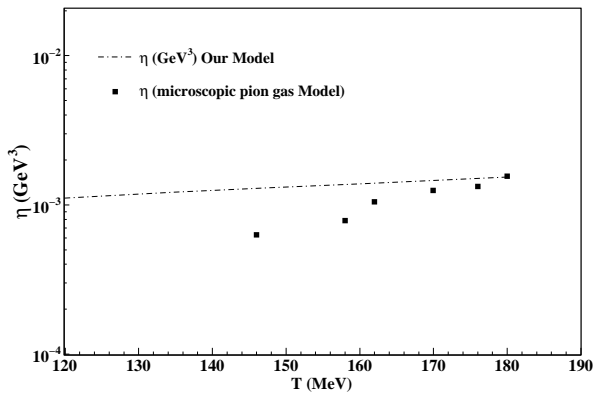}
\caption[]{Variation of $\eta$ with respect to temperature at $\mu_{B}=300 MeV$ in our model and a comparison with the results obtained by A. Muronga \cite{Muronga:2004}.}
\end{center}
\end{figure}

\begin{figure}
\begin{center}
\includegraphics[height=25em]{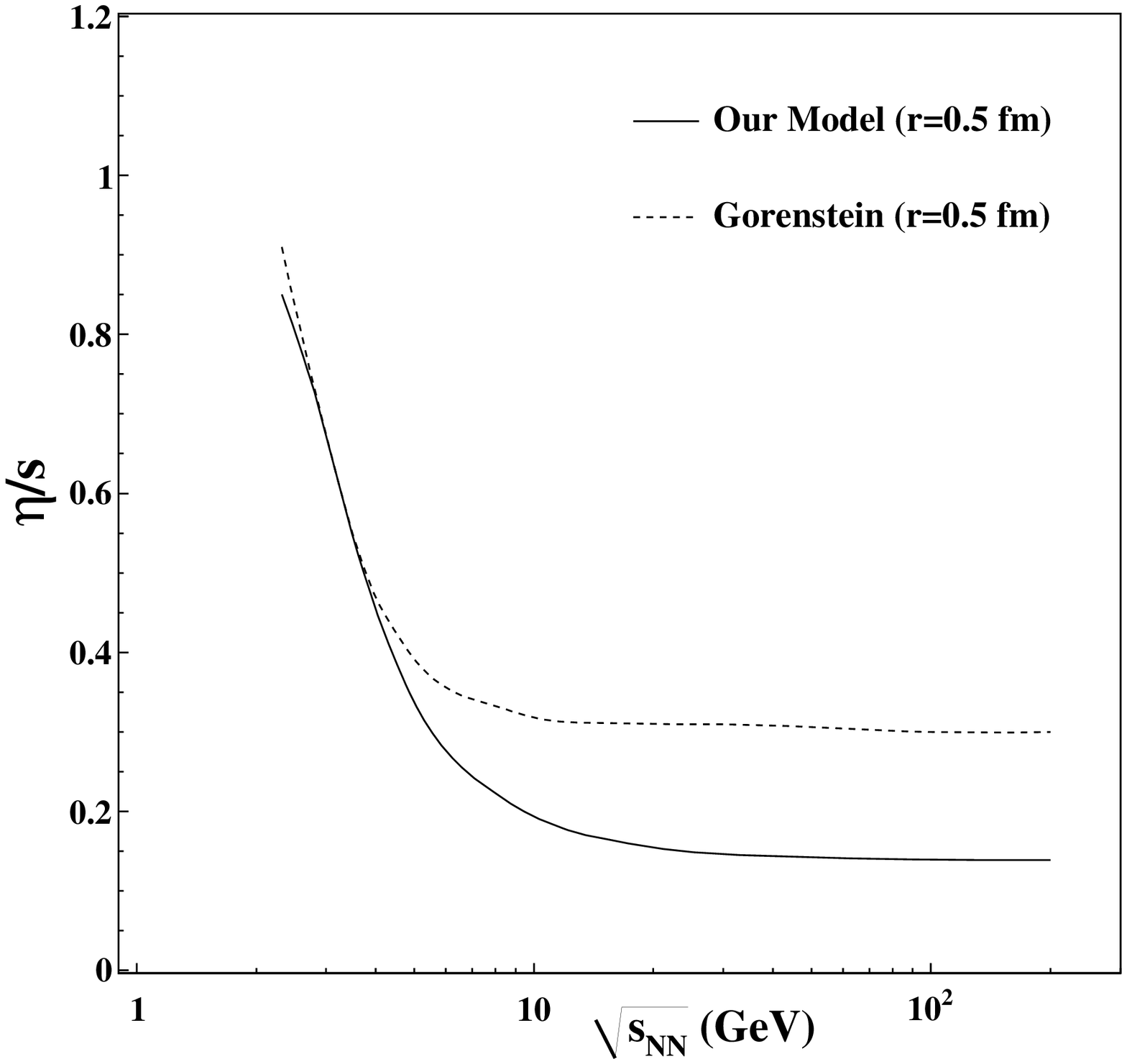}
\caption[]{Variation of $\eta/s$ with respect to $\sqrt{s_{NN}}$ in our model \cite{Tiwari:2012} and other model calculation \cite{Gorenstein:2008}.}
\end{center}
\end{figure}

In Fig. 16, we have shown the variation of $\eta/s$ with respect to $\mu_{B}$ at a very low temperature ($\approx 10$ MeV) \cite{Tiwari:2012}. Here we find that the $\eta/s$ is constant as $\mu_{B}$ increases upto $700$ MeV and then sharply decreases. This kind of valley-structure at low temperature and at $\mu_{B}$ around $950$ MeV was also obtained by J. W. Chen $et\; al.$ \cite{Chen:2007} and K. Itakura $et\; al.$ \cite{Itakura:2008}. They have related this structure to the liquid-gas phase transition of the nuclear matter. As we increase the temperature above $20$ MeV, this valley-like structure disappears. They further suspect that the observation of a discontinuity in the bottom of $\eta/s$ valley may correspond to the location of the critical point. Our HG model yields a curve in complete agreement with these results. Fig. 17 represents the variation of $\eta$ and $\eta/s$ with respect to temperature at a fixed $\mu_{B}$ ($=300$ MeV), for HG having a baryonic hard-core size $r=0.8$ fm. We have compared our result with the result obtained in Ref. \cite{Itakura:2008}. Here we find that $\eta$ increases with temperature in our HG model as well as in the simple phenomenological calculation \cite{Itakura:2008}, but decreases with increasing temperature in low temperature effective field theory (EFT) calculations \cite{Chen:2007,Itakura:2008}. However, $\eta/s$ decreases with increasing temperature in all three calculations and $\eta/s$ in our model gives the lowest value at all the  temperatures in comparison to other models. In Fig. 18, we have shown a comparison between $\eta$ calculated in our HG model with the results obtained in a microscopic pion gas model used in Ref. \cite{Muronga:2004}. Our model results show a fair agreement with the microscopic model results for the temperature higher than $160$ MeV while at lower temperatures the microscopic calculation predicts lower values of  $\eta$  in comparison to our results. The most probable reason may be that the calculations have been done only for pion gas in the microscopic model while at low temperatures the inclusion of baryons in the HG is very important in order to extract a correct value for the shear viscosity \cite{Tiwari:2012}. Figure 19 shows the variation of $\eta/s$ with respect to $\sqrt{s_{NN}}$ in our model calculation. We have compared our results with that calculated in \cite{Gorenstein:2008}. There is similarity in our results at lower energies while our results significantly differ at higher energies. Although both the calculations show that $\eta/s$ decreases with increasing $\sqrt{s_{NN}}$.

The study of the transport properties of non-equilibrium systems which are not far from an equilibrium state has yielded valuable results in the recent past. Large values of the elliptic flow observed at RHIC indicates that the matter in the fireball behaves as a nearly perfect liquid with a small value of the $\eta/s$ ratio. After evaluating $\eta/s$ in strongly coupled theories using AdS/CFT duality conjecture, a lower bound was reported as $\displaystyle\eta/s=\frac{1}{4\pi}$. We surprisingly notice that the fireball with hot, dense HG as described in our model gives transport coefficient which agree with those given in different approaches. Temperature and baryon chemical potential dependence of the $\eta/s$ are analyzed and compared with the results obtained in other models. Our results lend support to the claim that knowledge of the EOS and the transport coefficients of HG is essential for a better understanding of the dynamics of the medium formed in the heavy-ion collisions \cite{Tiwari:2012}.

\subsection{Rapidity and Transverse Mass Spectra}

In order to suggest any unambiguous signal for QGP, the dynamics of the collisions should be understood properly. Such information can be obtained by analyzing the properties of various particles which are emitted from various stages of the collisions. Hadrons are produced at the end of the hot and dense QGP phase, but they subsequently scatter in the confined hadronic phase prior to decoupling (or "freeze-out") from the collision system and finally a collective evolution of the hot and dense matter occurs in the form of transverse, radial or elliptic flow which are instrumental in shaping the important features of the particle spectra. The global properties and dynamics of freeze-out can be at best studied via hadronic observables such as rapidity distributions and transverse mass spectra etc. \cite{Letessier:2004}. There are various approaches for the study of rapidity as well as transverse mass spectra of HG \cite{Bjorken:1983,Braun:1995,Landau:1953,Schnedermann:1993,Schnedermann:1992,Feng:2011,Uddin:2012,Hirano:2002,Manninen:2011,Bass:1998,Mayer:1997,Becattini:2007,Biedron:2007,Cleymans:2008,Broniowski:2002,Adler:2001,Adcox:2004,Adcox:2002,Bozek:2008,Ivanov:2013}. Hadronic spectra from purely thermal models usually reveal an isotropic distribution of particles \cite{Huovinen:2002} and hence the rapidity spectra obtained with the purely thermal models do not reproduce the features of the experimental data satisfactorily. Similarly the transverse mass spectra from the thermal models reveal a more steeper curve than that observed experimentally. The comparisons illustrate that the fireball formed in heavy-ion collisions does not expand isotropically in nature and there is a prominent input of collective flow in the longitudinal and transverse directions which finally causes anisotropy in the rapidity and transverse mass distributions of the hadrons after the freeze-out. Here we mention some kinds of models of thermal and collective flow used in the literature. Hydrodynamical properties of the expanding fireball have been initially discussed by Bjorken and Landau for the central-rapidity and stopping regimes, respectively \cite{Bjorken:1983,Landau:1953}. However, collisions even at RHIC energies reveal that they are neither fully stopped, nor fully transparent. As the collision energy increases, the longitudinal flow grows stronger and leads to a cylindrical geometry as postulated in Ref. \cite{Braun:1995,Schnedermann:1993,Schnedermann:1992}. They assume that the fireballs are distributed uniformally in the longitudinal direction and demonstrate that the available data can consistently be described in a thermal model with inputs of chemical equilibrium and flow, although they have also used the experimental data for small systems only. They use two simple parameters : transverse flow velocity ($\beta_r$) and temperature ($T$) in their models. In Ref. \cite{Feng:2011}, non-uniform flow model is used to analyze the spectra specially to reproduce the dip at midrapidity in the rapidity spectra of baryons by assuming that the fireballs are distributed non-uniformly in the longitudinal phase space. In Ref. \cite{Becattini:2007,Biedron:2007,Cleymans:2008}, the rapidity-dependent baryon chemical potential has been invoked to study the rapidity spectra of hadrons. In certain hydrodynamical models \cite{Broniowski:2002}, measured transverse momentum ($p_T$) distributions in $Au-Au$ collisions at $\sqrt{s_{NN}}=130\;GeV$ \cite{Adler:2001,Adcox:2004,Adcox:2002} have been described successfully by incorporating a radial flow. In Ref. \cite{Bozek:2008}, rapidity spectra of mesons have been studied using viscous relativistic hydrodynamics in a 1+1 dimension assuming a non-boost invariant Bjorken flow in the longitudinal direction. They have also analyzed the effect of the shear viscosity on the longitudinal expansion of the matter. Shear viscosity counteracts the gradients of the velocity field, as a consequence slows down the longitudinal expansion. Ivanov \cite{Ivanov:2013} has employed 3FD model \cite{Ivanov:2006} for the study of rapidity distributions of hadrons in the energy range from $2.7\;GeV$ to $62.4\;GeV$. In 3 FD model, three different EOS : ($i$) a purely hadronic EOS ($ii$) the EOS involving first order phase transition from hot, dense HG to QGP and ($iii$) the EOS with smooth crossover transition are used. Within all three scenarios they reproduced the data at the almost same extent. In Ref. \cite{Mohanty:2003}, rapidity distributions of various hadrons in the central nucleus-nucleus collisions have been studied in the Landau's and Bjorken's hydrodynamical model. The effect of speed of sound ($c_s$) on the hadronic spectra and the correlation of $c_s$ with freeze-out parameters are indicated.

In this section, we study the rapidity and transverse mass spectra of hadrons using thermal approach. We can rewrite Eq. (42) in the following manner \cite{Tiwari:2013} : 

\begin{eqnarray}
\begin{aligned} 
n_i^{ex} 
=\frac{g_i\lambda_i}{(2\pi)^3}\;\Big[\Big((1-R)-\lambda_i\frac{\partial{R}}{\partial{\lambda_i}}\Big)\;\int_0^\infty \frac{d^3k}{\displaystyle \Big[exp\left(\frac{E_i}{T}\right)+\lambda_i \Big]}
\\
-\lambda_i(1-R)\;\int_0^\infty \frac{d^3k}{\displaystyle \Big[exp\left(\frac{E_i}{T}\right)+\lambda_i\Big]^2}\Big].
\end{aligned} 
\end{eqnarray}

It means that the invariant distributions are \cite{Braun:1995,Schnedermann:1993} :

\begin{eqnarray}
\begin{aligned} 
E_i\;\frac{d^3N_i}{dk^3}=\frac{g_iV\lambda_i}{(2\pi)^3}\;\Big[\Big((1-R)-\lambda_i\frac{\partial{R}}{\partial{\lambda_i}}\Big)\; \frac{E_i}{\displaystyle \Big[exp\left(\frac{E_i}{T}\right)+\lambda_i\Big]}\\
-\lambda_i(1-R)\;\frac{E_i}{\displaystyle \Big[exp\left(\frac{E_i}{T}\right)+\lambda_i\Big]^2}\Big].
\end{aligned}
\end{eqnarray}

If we use Boltzmann's approximation, Eq. (71) differs from the one used in the paper of Schnedermann $et\;al.$ \cite{Schnedermann:1993} by the presence of a prefactor $\displaystyle\Big[(1-R)-\lambda_i\frac{\partial{R}}{\partial{\lambda_i}}\Big]$. However, we measure all these quantities precisely at the chemical freeze-out using our model and hence quantitatively we do not require any normalizing factor as is required in Ref. \cite{Schnedermann:1993}. We use the following expression to calculate the rapidity distributions of baryons in the thermal model \cite{Tiwari:2013} : 

\begin{equation}
\begin{aligned} 
\Big(\frac{dN_i}{dy}\Big)_{th}=\frac{g_iV\lambda_i}{(2{\pi}^2)}\;\Big[\Big((1-R)-\lambda_i\frac{\partial{R}}{\partial{\lambda_i}}\Big)\; \int_0^\infty \frac{m_T^2\;coshy\;dm_T}{\displaystyle\Big[exp\left(\frac{m_T\;coshy}{T}\right)+\lambda_i\Big]}
\\
-\lambda_i(1-R)\;\int_0^\infty \frac{m_T^2\;coshy\;dm_T}{\displaystyle\Big[exp\left(\frac{m_T\;coshy}{T}\right)+\lambda_i\Big]^2}\Big].
\end{aligned} 
\end{equation}
Here $y$ is the rapidity variable and $m_T$ is the transverse mass $(m_T=\sqrt{{m}^2+{p_T}^2})$. Also $V$ is the total volume of the fireball formed at chemical freeze-out and $N_i$ is the total number of ith baryons. We assume that the freeze-out volume of the fireball for all types of hadrons at the time of the homogeneous emissions of hadrons remains the same. It can be mentioned here that in the above equation, there occurs no free parameter because all the quantities $g$, $V$, $\lambda$, $R$ etc. are determined in the model. Although, the above Eq. (72) describes the experimental data only at midrapidity while it fails at forward and backward rapidities, so we need to modify it by incorporating a flow factor in the longitudinal direction. Thus the resulting rapidity spectra of ith hadron is \cite{Braun:1995,Tiwari:2013,Schnedermann:1993}: 

\begin{eqnarray}
\frac{dN_i}{dy}=\int_{-\eta_{max.}}^{\eta_{max.}} \Big(\frac{dN_i}{dy}\Big)_{th}(y-\eta)\;d\eta,
\end{eqnarray}
where $\displaystyle\Big(\frac{dN_i}{dy}\Big)_{th}$ can be calculated by using Eq. (72). The expression for average longitudinal velocity is \cite{Feng:2011,Netrakanti:2005}:
 
\begin{eqnarray}
\langle\beta_L\rangle=tanh\Big(\frac{\eta_{max}}{2}\Big).
\end{eqnarray}
Here $\eta_{max}$ is a free parameter which provides the upper rapidity limit for the longitudinal flow velocity at a particular $\sqrt{s_{NN}}$ and it's value is determined by the best experimental fit. The value of $\eta_{max}$ increases with the increasing $\sqrt{s_{NN}}$ and hence $\beta_L$ also increases. Cleymans $et\;al.$ \cite{Cleymans:2008} have extended the thermal model \cite{Becattini:2007}, in which the chemical freeze-out parameters are rapidity-dependent, to calculate the rapidity spectra of hadrons. They use the following expression for rapidity spectra :
\begin{eqnarray}
\frac{dN^i}{dy}=\int_{-\infty}^{+\infty}\rho(y_{FB})\frac{dN_{1}^i(y-y_{FB})}{dy}dy_{FB},
\end{eqnarray}
where $\frac{dN_1^i}{dy}$ is the thermal rapidity distribution of particles calculated by using Eq. (72) and $\rho(y_{FB})$ is a Gaussian distribution of fireballs centered at zero and given by : 
\begin{eqnarray}
\rho(y_{FB})=\frac{1}{\sqrt{2\pi}\sigma}exp(\frac{-y^2_{FB}}{2\sigma^2}).
\end{eqnarray} 
Similarly we calculate the transverse mass spectra of hadrons by using following expression \cite{Tiwari:2013} :

\begin{eqnarray}
\frac{dN_i}{m_T\;dm_T}=\frac{g_iV\lambda_i}{(2{\pi}^2)}\;\Big[(1-R)-\lambda_i\frac{\partial{R}}{\partial{\lambda_i}}\Big]\;m_T\:K_1\Big(\frac{m_T}{T}\Big),
\end{eqnarray}

where $K_1\displaystyle\Big(\frac{m_T}{T}\Big)$ is the modified Bessel's function :

\begin{eqnarray}
K_1\Big(\frac{m_T}{T}\Big)=\int_0^\infty coshy\;\Big[exp\left(\frac{-m_T\;coshy}{T}\right)\Big]dy.
\end{eqnarray}
Above expression for transverse mass spectra arises from a stationary thermal source alone which is not capable of describing the experimental data successfully. So, we incorporate flow velocity in both the directions in the Eq. (77), longitudinal as well as transverse, in order to describe the experimental data satisfactorily. After defining the flow velocity field, we can calculate the invariant momentum spectrum by using the following formula \cite{Schnedermann:1993,Cooper:1974} :

\begin{eqnarray}
E_i\;\frac{d^3N_i}{dk^3}=\frac{g_iV\lambda_i}{(2\pi)^3}\;\Big[(1-R)-\lambda_i\frac{\partial{R}}{\partial{\lambda_i}}\Big]\int exp\Big(\frac{-k_{\mu}u^{\mu}}{T}\Big)\;k_\lambda\;d\sigma_\lambda.
\end{eqnarray}
While deriving Eq.(79), we assume that the local fluid velocity $u^{\mu}$ give a boost to an isotropic thermal distribution of hadrons. Now the final expression of transverse mass spectra of hadrons after incorporation of flow velocity in our model is \cite{Tiwari:2013} :

\begin{eqnarray}
\frac{dN_i}{m_Tdm_T}=\frac{g_iV\lambda_i\;m_T}{(2{\pi}^2)}\;\Big[(1-R)-\lambda_i\frac{\partial{R}}{\partial{\lambda_i}}\Big]\int_0^{R_{0}} r\;dr\;K_1\Big(\frac{m_T\;cosh\rho}{T}\Big)I_0\Big(\frac{p_T\;sinh\rho}{T}\Big).
\end{eqnarray}
Here $I_0\displaystyle\Big(\frac{p_T\;sinh\rho}{T}\Big)$ is the modified Bessel's function :

\begin{eqnarray}
I_0\Big(\frac{p_T\;sinh\rho}{T}\Big)=\frac{1}{2\pi}\int_0^{2\pi} exp\Big(\frac{p_T\;sinh\rho\;cos\phi}{T}\Big)d\phi,
\end{eqnarray}
where $\rho$ is given by $\rho=tanh^{-1}\beta_r$, with the velocity profile chosen as $\beta_r=\displaystyle\beta_s\;\Big(\xi\Big)^n$ \cite{Braun:1995,Schnedermann:1993}. $\beta_s$ is the maximum surface velocity and is treated as a free parameter and $\xi=\displaystyle\Big(r/R_0\Big)$. The average of the transverse velocity can be evaluated as \cite{Adcox:2004} :

\begin{eqnarray}
<\beta_r> =\frac{\int \beta_s\xi^n\xi\;d\xi}{\int \xi\;d\xi}=\Big(\frac{2}{2+n}\Big)\beta_s.
\end{eqnarray}

In our calculation we use a linear velocity profile, ($n=1$) and $R_0$ is the maximum radius of the expanding source at freeze-out ($0<\xi<1$) \cite{Adcox:2004}.

\begin{figure}
\includegraphics[height=22em]{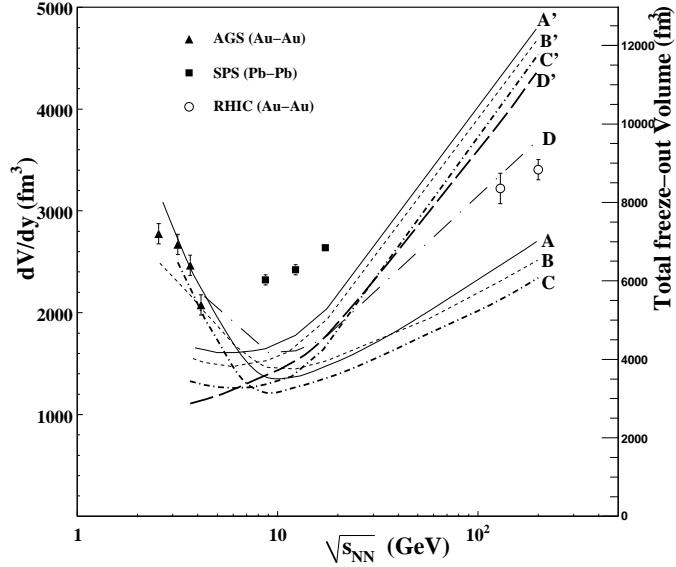}
\caption[]{ Energy dependence of the freeze-out volume for the central nucleus-nucleus collisions. The symbols are the HBT data for freeze-out volume $V_{HBT}$ for the $\pi^+$ \cite{Adamova:2003}. $A'$ ,$B'$ and $C'$ are the total freeze-out volume and $A$ ,$B$ and $C$ depict the $dV/dy$ as found in our model for $\pi^+$, $K^+$ and $K^-$ , respectively. $D$ represents the total freeze-out volume for $\pi^+$ calculated in the Ideal HG model. $D'$ is the the total freeze-out volume for $\pi^+$ in our model calculation using Boltzmann's statistics \cite{Tiwari:2013}.}
\end{figure}

\begin{figure}
\includegraphics[height=20em]{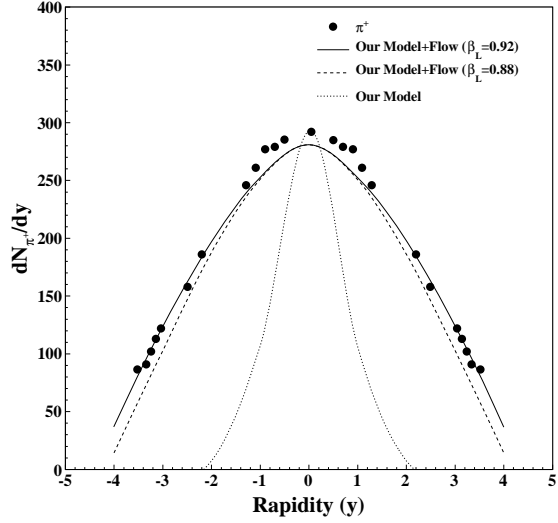}
\caption[]{Rapidity distribution of $\pi^+$ at $\sqrt{s_{NN}}= 200 GeV $. Dotted line shows the rapidity distribution calculated in our thermal model \cite{Tiwari:2013}. Solid line and dashed line show the results obtained after incorporating longitudinal flow in our thermal model. Symbols are the experimental data \cite{Bearden1:2005}.}
\end{figure}

\begin{figure}
\includegraphics[height=20em]{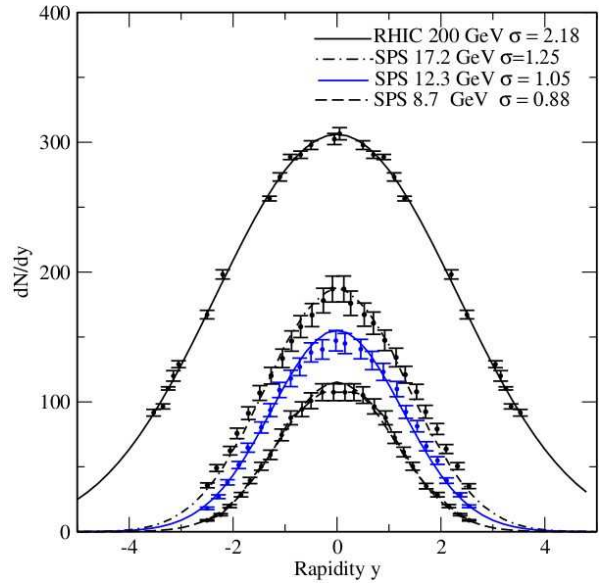}
\caption[]{Rapidity distribution of pion at various $\sqrt{s_{NN}}$. Lines are model results and points are experimental data. Figure is taken from Ref. \cite{Cleymans:2008}.}
\end{figure}

In Fig. 20, we have shown the variations of $V$ and $dV/dy$ with the $\sqrt{s_{NN}}$ calculated in our excluded-volume model and compared with the results of various thermal models. We show the total freeze-out volume for $\pi^+$ calculated in our model using Boltzmann's statistics. We see that there is a significant difference between the results arising from quantum statistics and Boltzmann's statistics \cite{Tiwari:2013}. We also show the total freeze-out volume for $\pi^+$ in IHG model calculation by dash-dotted line $D$. We clearly notice a remarkable difference between the results of our excluded-volume model and that of IHG model also. We have also compared predictions from our model with the data obtained from the pion interferometry (HBT) \cite{Adamova:2003} which in fact reveals thermal (kinetic) freeze-out volumes. The results of thermal models support the finding that the decoupling of strange mesons from the fireball takes place earlier than the $\pi$-mesons. Moreover, a flat minimum occurs in the curves around the center-of-mass energy $\approx8\;GeV$ and this feature is well supported by HBT data. In Fig. 21, we present the rapidity distribution of $\pi^+$ for central $Au+Au$ collisions at $\sqrt{s_{NN}}=200\; GeV$ over full rapidity range. Dotted line shows the distribution of $\pi^+$ due to purely thermal model. Solid line shows the rapidity distributions of $\pi^+$ after the incorporation of longitudinal flow in our thermal model and results give a good agreement with the experimental data \cite{Bearden1:2005}. In fitting the experimental data, we use the value of $\eta_{max}=3.2$ and hence the longitudinal flow velocity $\beta_L=0.92$ at $\sqrt{s_{NN}}=200\; GeV$. For comparison and testing the appropriateness of this parameter, we also show the rapidity distributions at a different value i.e., $\eta_{max}=2.8$ (or, $\beta_L=0.88$), by a dashed line in the figure. We find that the results slightly differ and hence it shows a small dependence on $\eta_{max}$ \cite{Tiwari:2013}. Fig. 22 represents the rapidity distributions of pion at various $\sqrt{s_{NN}}$ calculated by using Eq. (75) \cite{Cleymans:2008}. There is a good agreement between the model results and experimental data at all $\sqrt{s_{NN}}$.

\begin{figure}
\includegraphics[height=20em]{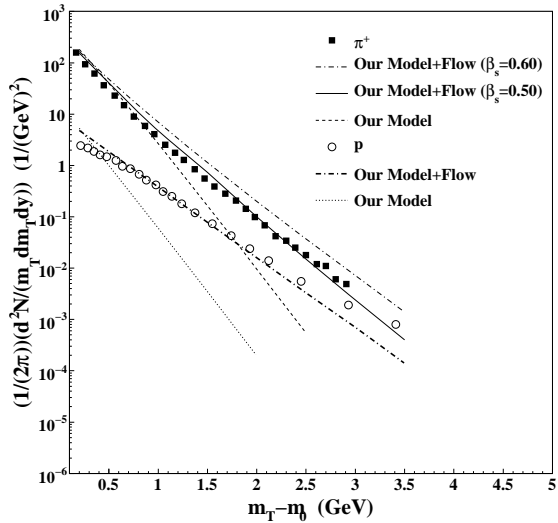}
\caption[]{Transverse mass spectra for $\pi^+$ and proton for the most central collisions at $\sqrt{s_{NN}}= 200\; GeV$. Dashed and dotted lines are the transverse mass spectra due to purely thermal source for $\pi^+$ and proton, respectively. Solid and dash-dotted lines are the results for $\pi^+$ and proton, respectively obtained after incorporation of flow in thermal model \cite{Tiwari:2013}. Symbols are the experimental data \cite{Adler:2004}.}
\end{figure}

\begin{figure}
\includegraphics[height=20em]{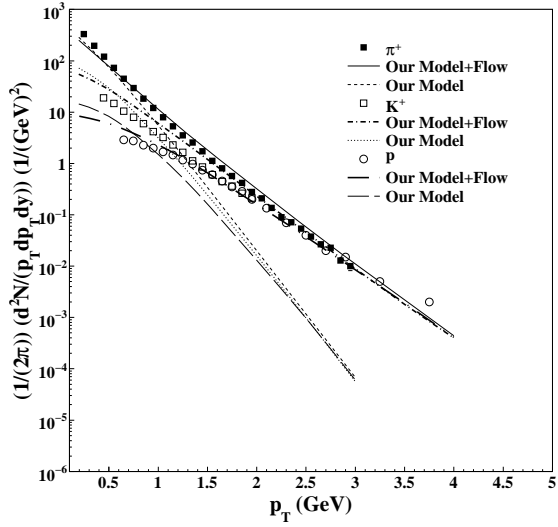}
\caption[]{Transverse momentum spectra for $\pi^+$, $p$, and $K^+$ for the most central $Au-Au$ collision at $\sqrt{s_{NN}}=200\:GeV$ \cite{Tiwari:2013}. Lines are the results of our model calculation and symbols are the experimental results \cite{Adler:2004}.}
\end{figure}

\begin{figure}
\includegraphics[height=22em]{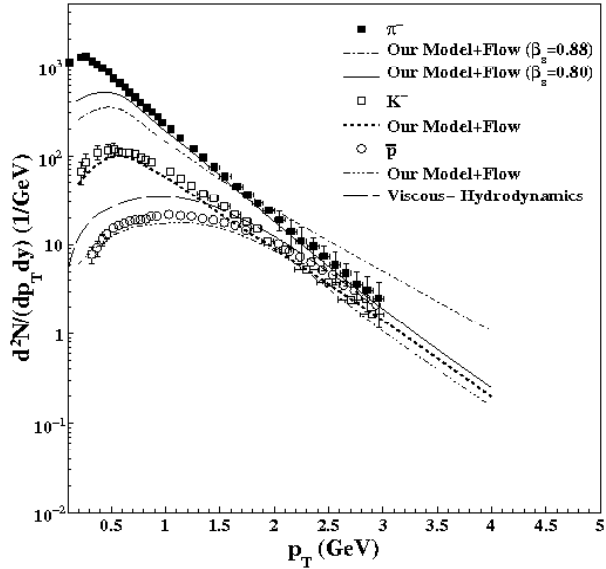}
\caption[]{Transverse momentum spectra of various hadrons for the most central collisions of $Pb-Pb$ at $\sqrt{s_{NN}}=2.76\;TeV$ from LHC \cite{Tiwari:2013}. Lines are the results of model calculations and symbols are the experimental results \cite{Floris:2011}. Thick-dashed line is the prediction of viscous-hydrodynamical model \cite{Shen:2011} for $\bar{p}$.}
\end{figure}

In Fig. 23, we show the transverse mass spectra for $\pi^+$ and proton for the most central collisions of $Au+Au$ at $\sqrt{s_{NN}}= 200\;GeV$. We have neglected the contributions from the resonance decays in our calculations since these contributions affect the transverse mass spectra only towards the lower transverse mass side i.e. $m_T<0.3 \;GeV$. We find a good agreement between our calculations and the experimental results for all $m_T$ except $m_T<0.3 \;GeV$ after incorporating the flow velocity in purely thermal model. This again shows the importance of collective flow in the description of the experimental data \cite{Adler:2004}. At this energy, the value of $\beta_s$ is taken as $0.50$ and transverse flow velocity $\beta_r=0.33$. This set of transverse flow velocity is able to reproduce the transverse mass spectra of almost all the hadrons at $\sqrt{s_{NN}}= 200\; GeV$. We notice that the transverse flow velocity slowly increases with the increasing $\sqrt{s_{NN}}$. If we take $\beta_s=0.60$, we find that the results differ with data as shown in Fig. 23. In Fig. 24, we show the transverse momentum ($p_{T}$) spectra for $\pi^{+}$, $K^+$ and $p$ in the most central collisions of $Au-Au$ at $\sqrt{s_{NN}}=200\;GeV$. Our model calculations reveal a close agreement with the experimental data \cite{Adler:2004}. In Fig. 25, we show the $p_{T}$ spectra of $\pi^{-}$, $K^-$ and $\bar{p}$ for the $Pb-Pb$ collisions at $\sqrt{s_{NN}}=2.76\;TeV$ at the LHC. Our calculations again give a good fit to the experimental results \cite{Floris:2011}. We also compare our results for $\bar{p}$ spectrum with the hydrodynamical model of Shen $et\; al.$ \cite{Shen:2011}, which successfully explains $\pi^{-}$, and $K^-$ spectra but strongly fails in the case of $\bar{p}$ spectrum \cite{Tiwari:2013}. In comparison, our model results show closer agreement with the experimental data. Shen $et\; al.$ \cite{Shen:2011} have employed $(2+1)$-dimensional viscous hydrodynamics with the lattice QCD-based EOS. They use Cooper-Frye prescription to implement kinetic freeze-out in converting the hydrodynamic output into the particle spectra. Due to lack of a proper theoretical and phenomenological knowledge, they use the same parameters for $Pb-Pb$ collisions at LHC energy, which was used for $Au-Au$ collisions at $\sqrt{s_{NN}}=200\;GeV$. Furthermore, they use the temperature independent $\eta/s$ ratio in their calculation. After fitting the experimental data, we get $\beta_{s}=0.80$ $(\beta_{r}=0.53)$ at this energy which indicates the collective flow becoming stronger at LHC energy than that observed at RHIC energies. In this plot, we also attempt to show how the spectra for $\pi^-$ will change at a slightly different value of the parameter $i. e.$, $\beta_s=0.88$ \cite{Tiwari:2013}.

\section{Summary and Conclusions}

The main aim in this article is to emphasize the use of the thermal approach in describing various yields of different particle species that have been measured in various experiments running at various places. We have discussed various types of thermal approaches for the formulation of EOS for HG. We have argued that, incorporation of interactions between hadrons in a thermodynamically consistent way is important for the realistic formulation of HG from both qualitatively and quantitatively point of view. We have presented the systematic study of the particle production in heavy-ion collisions from AGS to LHC energy. We have observed from this analysis that the production of the particles seems to be occurred according to principle of equilibrium. Yields of hadrons and their ratios measured in heavy-ion collisions match with the predictions of thermal models assured the thermalization of the collision fireball formed in heavy-ion collisions. Furthermore, various experimental observables such as transverse momentum spectra, elliptic flow etc. indicate the presence of the thermodynamical pressure developed in the early stage, and correlations which are expected in a thermalized medium.

We have discussed a detailed formulation of various excluded-volume models and their shortcomings. Some excluded-volume models are not thermodynamically consistent because they do not possess a well defined partition function from which various thermodynamical quantities such as number density etc. can be calculated. However, some of them are the thermodynamically consistent but suffer from some unphysical situations cropping up in the calculations. We have proposed a new approximately thermodynamically consistent excluded-volume model for a hot and dense HG. We have used quantum statistics in the grand canonical partition function of our model so that it works even at extreme values of $T$ and $\mu_B$ where all other models fail. Moreover, our model respects causality. We have presented the calculations of various thermodynamical quantities such as entropy per baryon and energy density etc. in various excluded-volume models and compare the results with that of a microscopic approach URASiMA. We find that our model results are in close agreement with that of the entirely different approach URASiMA model. We have calculated various particle ratios at various $\sqrt{s_{NN}}$ and confronted the results of various thermal models with the experimental data and find that they are indeed successful in describing the particle ratios. Although, we find that our model results are closer to the experimental data in comparison to that of other excluded-volume models. We have calculated some conditions such as $E/N$, $S/N$ etc. at chemical freeze-out points and attempted to test whether these conditions involve energy independence as well as independence of structure of the nuclei involved in the collisions. We find that $E/N\approx1.0\; GeV$, and $S/N\approx7.0$ are the two robust freeze-out criteria which show independence of the energy and structure of nuclei. Moreover, the calculations of transport properties in our model match well with the results obtained in other widely different approaches. Further, we present an analysis of rapidity distributions and transverse mass spectra of hadrons in central nucleus-nucleus collision at various $\sqrt{s_{NN}}$ using our EOS for HG. We see that the stationary thermal source alone cannot describe the experimental data fully unless we incorporate flow velocities in the longitudinal as well as in the transverse direction and as a result, our modified model predictions show a good agreement with the experimental data. Our analysis shows that a collective flow develops at each $\sqrt{s_{NN}}$ which increases further with the increasing $\sqrt{s_{NN}}$. The description of the rapidity distributions and transverse mass spectra of hadrons at each $\sqrt{s_{NN}}$ matches very well with the experimental data. Thus, we emphasize that thermal models are indeed an important tool to describe the various features of hot and dense HG. Although, these models are not capable of telling whether QGP was formed before HG phase but can give an indirect indication of it by showing any anomalous feature as observed in the experimental data.

In conclusion, the net outcome of this review is indeed a surprising one. The excluded-volume HG models are really successfull in describing all kinds of features of the HG formed in ultra-relativistic heavy-ion collisions. Most important property indicated by such description is the chemical equilibrium reached in such collisions. However, the description is still a geometrical one and does not involve any microscopic picture of interactions. Moreover, it's relativistic and field theoretic generalizations are still needed in order to make the picture a more realistic description. But it is amazing to find that these models still work much better than expected. Notwithstanding these remarks, we should add that Lattice QCD results are now available for the pressure, entropy density and energy density etc. for the entire temperature range from $T=0$ to higher values at $\mu=0$. Here low temperature phase of QCD is the HG and recently our excluded-volume model reproduces these properties quite in agreement with Lattice results \cite{Srivastava:2012}. We have also used these calculations in the precise determination of the QCD critical end point \cite{Srivastava:2010}. Thus we conclude that the excluded-volume models are successfull in reproducing numerical results obtained in various experiments and, therefore, further research is required to show how these descriptions are connected with the microscopic interactions.

\section{Acknowledgments}
SKT is grateful to Council of Scientific and Industrial Research (CSIR), New Delhi for providing a research grant.

\end{document}